\documentclass{eptcs}
\providecommand{\event}{TFPIE 2025}

\usepackage{subcaption}
\usepackage{amsmath}
\usepackage{bm}
\usepackage{hyperref}
\usepackage{cleveref}
\usepackage{alltt}
\usepackage{tikz}
\usepackage{float}
\floatstyle{ruled}
\restylefloat{figure}
\usepackage{pgfplots}
\pgfplotsset{compat=1.18}
\usepackage{balance}
\usepackage{listings}
\usepackage{calc}
\usepackage{enumitem}
\usepackage[T1]{fontenc}

	\lstdefinelanguage{racket} {
		morecomment=[l]{;},         % Inline comments start with ;
		morecomment=[s]{\#|}{|\#},  % Block comments are done with #|  |#
		commentstyle={\color{darkorange}\slshape\sffamily},
		morestring=[b]",
		stringstyle=\color{red},
		literate=%
		{->}{{$\rightarrow$}}1
		{*}{{*}}1
		{lambda}{{\textcolor{blue}{\(\lambda\)}}}1
		{EMP}{{\ep{}}}1,
		classoffset=1,
		morekeywords={
			define, define-syntax, define-macro, define-datatype, define-stream, \#lang, stream-lambda
		},
		keywordstyle=\color{purple},
		classoffset=2,
		morekeywords={
			*, *list/c, +, -, /, <, </c, <=, =, >, >/c, >=, abort-current-continuation, abs, absolute-path?, acos, add1, alarm-evt, always-evt, andmap, angle, append, append*, append-map, argmax, argmin, arithmetic-shift, arity-at-least-value, arity-at-least?, arity-checking-wrapper, arity-includes?, arity=?, arrow-contract-info-accepts-arglist, arrow-contract-info-chaperone-procedure, arrow-contract-info-check-first-order, arrow-contract-info?, asin, assert-unreachable, assf, assoc, assq, assv, assw, atan, banner, base->-doms/c, base->-rngs/c, base->?, bitwise-and, bitwise-bit-field, bitwise-bit-set?, bitwise-ior, bitwise-not, bitwise-xor, blame-add-car-context, blame-add-cdr-context, blame-add-missing-party, blame-add-nth-arg-context, blame-add-range-context, blame-add-unknown-context, blame-context, blame-contract, blame-fmt->-string, blame-missing-party?, blame-negative, blame-original?, blame-positive, blame-replace-negative, blame-replaced-negative?, blame-source, blame-swap, blame-swapped?, blame-update, blame-value, blame?, block-device-type-bits, boolean=?, boolean?, bound-identifier=?, box, box-cas!, box-immutable, box?, break-enabled, break-parameterization?, break-thread, build-chaperone-contract-property, build-compound-type-name, build-contract-property, build-flat-contract-property, build-list, build-path, build-path/convention-type, build-string, build-vector, byte-pregexp, byte-pregexp?, byte-ready?, byte-regexp, byte-regexp?, byte?, bytes, bytes->immutable-bytes, bytes->list, bytes->path, bytes->path-element, bytes->string/latin-1, bytes->string/locale, bytes->string/utf-8, bytes-append, bytes-append*, bytes-close-converter, bytes-convert, bytes-convert-end, bytes-converter?, bytes-copy, bytes-copy!, bytes-environment-variable-name?, bytes-fill!, bytes-join, bytes-length, bytes-no-nuls?, bytes-open-converter, bytes-ref, bytes-set!, bytes-utf-8-index, bytes-utf-8-length, bytes-utf-8-ref, bytes<?, bytes=?, bytes>?, bytes?, caaaar, caaadr, caaar, caadar, caaddr, caadr, caar, cadaar, cadadr, cadar, caddar, cadddr, caddr, cadr, call-in-continuation, call-in-nested-thread, call-with-break-parameterization, call-with-composable-continuation, call-with-continuation-barrier, call-with-continuation-prompt, call-with-current-continuation, call-with-default-reading-parameterization, call-with-escape-continuation, call-with-exception-handler, call-with-immediate-continuation-mark, call-with-input-bytes, call-with-input-string, call-with-output-bytes, call-with-output-string, call-with-parameterization, call-with-semaphore, call-with-semaphore/enable-break, call-with-values, call/cc, call/ec, car, cartesian-product, cdaaar, cdaadr, cdaar, cdadar, cdaddr, cdadr, cdar, cddaar, cddadr, cddar, cdddar, cddddr, cdddr, cddr, cdr, ceiling, channel-get, channel-put, channel-put-evt, channel-put-evt?, channel-try-get, channel?, chaperone-box, chaperone-channel, chaperone-continuation-mark-key, chaperone-contract-property?, chaperone-contract?, chaperone-evt, chaperone-hash, chaperone-hash-set, chaperone-of?, chaperone-procedure, chaperone-procedure*, chaperone-prompt-tag, chaperone-struct, chaperone-struct-type, chaperone-vector, chaperone-vector*, chaperone?, char->integer, char-alphabetic?, char-blank?, char-ci<=?, char-ci<?, char-ci=?, char-ci>=?, char-ci>?, char-downcase, char-extended-pictographic?, char-foldcase, char-general-category, char-grapheme-break-property, char-grapheme-step, char-graphic?, char-in, char-iso-control?, char-lower-case?, char-numeric?, char-punctuation?, char-ready?, char-symbolic?, char-title-case?, char-titlecase, char-upcase, char-upper-case?, char-utf-8-length, char-whitespace?, char<=?, char<?, char=?, char>=?, char>?, char?, character-device-type-bits, check-duplicate-identifier, checked-procedure-check-and-extract, choice-evt, class->interface, class-info, class-seal, class-unseal, class?, cleanse-path, close-input-port, close-output-port, coerce-chaperone-contract, coerce-chaperone-contracts, coerce-contract, coerce-contract/f, coerce-contracts, coerce-flat-contract, coerce-flat-contracts, collect-garbage, collection-file-path, collection-path, combinations, combine-output, compile, compile-allow-set!-undefined, compile-context-preservation-enabled, compile-enforce-module-constants, compile-syntax, compile-target-machine?, compiled-expression-add-target-machine, compiled-expression-recompile, compiled-expression?, compiled-module-expression?, complete-path?, complex/c, complex?, compose, compose1, conjoin, conjugate, cons, cons?, const, const*, continuation-mark-key?, continuation-mark-set->context, continuation-mark-set->iterator, continuation-mark-set->list, continuation-mark-set->list*, continuation-mark-set-first, continuation-mark-set?, continuation-marks, continuation-prompt-available?, continuation-prompt-tag?, continuation?, contract-continuation-mark-key, contract-custom-write-property-proc, contract-equivalent?, contract-first-order, contract-first-order-passes?, contract-late-neg-projection, contract-name, contract-proc, contract-projection, contract-property?, contract-random-generate, contract-random-generate-env?, contract-random-generate-fail, contract-random-generate-fail?, contract-random-generate-get-current-environment, contract-random-generate-stash, contract-random-generate/choose, contract-stronger?, contract-struct-exercise, contract-struct-generate, contract-struct-late-neg-projection, contract-struct-list-contract?, contract-val-first-projection, contract?, convert-stream, copy-file, copy-port, cos, cosh, count, current-blame-format, current-break-parameterization, current-code-inspector, current-command-line-arguments, current-compile, current-compile-realm, current-compile-target-machine, current-compiled-file-roots, current-continuation-marks, current-custodian, current-directory, current-directory-for-user, current-drive, current-environment-variables, current-error-message-adjuster, current-error-port, current-eval, current-evt-pseudo-random-generator, current-force-delete-permissions, current-future, current-gc-milliseconds, current-get-interaction-evt, current-get-interaction-input-port, current-inexact-milliseconds, current-inexact-monotonic-milliseconds, current-input-port, current-inspector, current-library-collection-links, current-library-collection-paths, current-load, current-load-extension, current-load-relative-directory, current-load/use-compiled, current-locale, current-logger, current-memory-use, current-milliseconds, current-module-declare-name, current-module-declare-source, current-module-name-resolver, current-module-path-for-load, current-namespace, current-output-port, current-parameterization, current-plumber, current-preserved-thread-cell-values, current-print, current-process-milliseconds, current-prompt-read, current-pseudo-random-generator, current-read-interaction, current-reader-guard, current-readtable, current-seconds, current-security-guard, current-subprocess-custodian-mode, current-subprocess-keep-file-descriptors, current-thread, current-thread-group, current-thread-initial-stack-size, current-write-relative-directory, curry, curryr, custodian-box-value, custodian-box?, custodian-limit-memory, custodian-managed-list, custodian-memory-accounting-available?, custodian-require-memory, custodian-shut-down?, custodian-shutdown-all, custodian?, custom-print-quotable-accessor, custom-print-quotable?, custom-write-accessor, custom-write-property-proc, custom-write?, date*-nanosecond, date*-time-zone-name, date*?, date-day, date-dst?, date-hour, date-minute, date-month, date-second, date-time-zone-offset, date-week-day, date-year, date-year-day, date?, datum->syntax, datum-intern-literal, default-continuation-prompt-tag, default-global-port-print-handler, degrees->radians, delete-directory, delete-file, denominator, dict-iter-contract, dict-key-contract, dict-value-contract, directory-exists?, directory-list, directory-type-bits, disjoin, display, displayln, double-flonum?, drop, drop-common-prefix, drop-right, dropf, dropf-right, dump-memory-stats, dup-input-port, dup-output-port, dynamic-get-field, dynamic-object/c, dynamic-require, dynamic-require-for-syntax, dynamic-send, dynamic-set-field!, dynamic-wind, eighth, empty, empty-sequence, empty-stream, empty?, environment-variables-copy, environment-variables-names, environment-variables-ref, environment-variables-set!, environment-variables?, eof, eof-object?, ephemeron-value, ephemeron?, eprintf, eq-contract-val, eq-contract?, eq-hash-code, eq?, equal-always-hash-code, equal-always-hash-code/recur, equal-always-secondary-hash-code, equal-always?, equal-always?/recur, equal-contract-val, equal-contract?, equal-hash-code, equal-hash-code/recur, equal-secondary-hash-code, equal<\%>, equal?, equal?/recur, eqv-hash-code, eqv?, error, error-contract->adjusted-string, error-display-handler, error-escape-handler, error-message->adjusted-string, error-message-adjuster-key, error-print-context-length, error-print-source-location, error-print-width, error-syntax->string-handler, error-value->string-handler, eval, eval-jit-enabled, eval-syntax, even?, evt/c, evt?, exact->inexact, exact-ceiling, exact-floor, exact-integer?, exact-nonnegative-integer?, exact-positive-integer?, exact-round, exact-truncate, exact?, executable-yield-handler, exit, exit-handler, exn-continuation-marks, exn-message, exn:break-continuation, exn:break:hang-up?, exn:break:terminate?, exn:break?, exn:fail:contract:arity?, exn:fail:contract:blame-object, exn:fail:contract:blame?, exn:fail:contract:continuation?, exn:fail:contract:divide-by-zero?, exn:fail:contract:non-fixnum-result?, exn:fail:contract:variable-id, exn:fail:contract:variable?, exn:fail:contract?, exn:fail:filesystem:errno-errno, exn:fail:filesystem:errno?, exn:fail:filesystem:exists?, exn:fail:filesystem:missing-module-path, exn:fail:filesystem:missing-module?, exn:fail:filesystem:version?, exn:fail:filesystem?, exn:fail:network:errno-errno, exn:fail:network:errno?, exn:fail:network?, exn:fail:object?, exn:fail:out-of-memory?, exn:fail:read-srclocs, exn:fail:read:eof?, exn:fail:read:non-char?, exn:fail:read?, exn:fail:syntax-exprs, exn:fail:syntax:missing-module-path, exn:fail:syntax:missing-module?, exn:fail:syntax:unbound?, exn:fail:syntax?, exn:fail:unsupported?, exn:fail:user?, exn:fail?, exn:misc:match?, exn:missing-module-accessor, exn:missing-module?, exn:srclocs-accessor, exn:srclocs?, exn?, exp, expand, expand-once, expand-syntax, expand-syntax-once, expand-syntax-to-top-form, expand-to-top-form, expand-user-path, explode-path, expt, externalizable<\%>, failure-result/c, false, false/c, false?, field-names, fifo-type-bits, fifth, file-exists?, file-name-from-path, file-or-directory-identity, file-or-directory-modify-seconds, file-or-directory-permissions, file-or-directory-stat, file-or-directory-type, file-position, file-position*, file-size, file-stream-buffer-mode, file-stream-port?, file-truncate, file-type-bits, filename-extension, filesystem-change-evt, filesystem-change-evt-cancel, filesystem-change-evt?, filesystem-root-list, filter, filter-map, filter-not, filter-read-input-port, find-compiled-file-roots, find-executable-path, find-library-collection-links, find-library-collection-paths, find-system-path, findf, first, fixnum?, flat-contract, flat-contract-predicate, flat-contract-property?, flat-contract?, flat-named-contract, flatten, floating-point-bytes->real, flonum?, floor, flush-output, fold-files, foldl, foldr, for-each, force, format, fourth, fprintf, free-identifier=?, free-label-identifier=?, free-template-identifier=?, free-transformer-identifier=?, fsemaphore-count, fsemaphore-post, fsemaphore-try-wait?, fsemaphore-wait, fsemaphore?, future, future?, futures-enabled?, gcd, generate-member-key, generate-temporaries, generic-set?, generic?, gensym, get-output-bytes, get-output-string, get/build-late-neg-projection, get/build-val-first-projection, getenv, global-port-print-handler, group-by, group-execute-bit, group-permission-bits, group-read-bit, group-write-bit, guard-evt, handle-evt, handle-evt?, has-blame?, has-contract?, hash, hash->list, hash-clear, hash-clear!, hash-copy, hash-count, hash-empty?, hash-ephemeron?, hash-eq?, hash-equal-always?, hash-equal?, hash-eqv?, hash-for-each, hash-has-key?, hash-iterate-first, hash-iterate-key, hash-iterate-key+value, hash-iterate-next, hash-iterate-pair, hash-iterate-value, hash-keys, hash-keys-subset?, hash-map, hash-placeholder?, hash-ref, hash-ref!, hash-ref-key, hash-remove, hash-remove!, hash-set, hash-set!, hash-set*, hash-set*!, hash-strong?, hash-update, hash-update!, hash-values, hash-weak?, hash?, hashalw, hasheq, hasheqv, identifier-binding, identifier-binding-portal-syntax, identifier-binding-symbol, identifier-distinct-binding, identifier-label-binding, identifier-prune-lexical-context, identifier-prune-to-source-module, identifier-remove-from-definition-context, identifier-template-binding, identifier-transformer-binding, identifier?, identity, if/c, imag-part, immutable?, impersonate-box, impersonate-channel, impersonate-continuation-mark-key, impersonate-hash, impersonate-hash-set, impersonate-procedure, impersonate-procedure*, impersonate-prompt-tag, impersonate-struct, impersonate-vector, impersonate-vector*, impersonator-contract?, impersonator-ephemeron, impersonator-of?, impersonator-prop:application-mark, impersonator-prop:blame, impersonator-prop:contracted, impersonator-property-accessor-procedure?, impersonator-property?, impersonator?, implementation?, implementation?/c, in-combinations, in-cycle, in-dict-pairs, in-parallel, in-permutations, in-sequences, in-values*-sequence, in-values-sequence, index-of, index-where, indexes-of, indexes-where, inexact->exact, inexact-real?, inexact?, infinite?, input-port?, inspector-superior?, inspector?, integer->char, integer->integer-bytes, integer-bytes->integer, integer-length, integer-sqrt, integer-sqrt/remainder, integer?, interface->method-names, interface-extension?, interface?, internal-definition-context-add-scopes, internal-definition-context-binding-identifiers, internal-definition-context-introduce, internal-definition-context-seal, internal-definition-context-splice-binding-identifier, internal-definition-context?, is-a?, is-a?/c, keyword->string, keyword-apply, keyword-apply/dict, keyword<?, keyword?, keywords-match, kill-thread, last, last-pair, lcm, length, liberal-define-context?, link-exists?, list, list*, list->bytes, list->mutable-set, list->mutable-setalw, list->mutable-seteq, list->mutable-seteqv, list->set, list->setalw, list->seteq, list->seteqv, list->string, list->vector, list->weak-set, list->weak-setalw, list->weak-seteq, list->weak-seteqv, list-contract?, list-prefix?, list-ref, list-set, list-tail, list-update, list?, listen-port-number?, load, load-extension, load-on-demand-enabled, load-relative, load-relative-extension, load/cd, load/use-compiled, local-expand, local-expand/capture-lifts, local-transformer-expand, local-transformer-expand/capture-lifts, locale-string-encoding, log, log-all-levels, log-level-evt, log-level?, log-max-level, log-message, log-receiver?, logger-name, logger?, magnitude, make-arity-at-least, make-base-empty-namespace, make-base-namespace, make-bytes, make-channel, make-chaperone-contract, make-continuation-mark-key, make-continuation-prompt-tag, make-contract, make-custodian, make-custodian-box, make-date, make-date*, make-derived-parameter, make-directory, make-directory*, make-do-sequence, make-empty-namespace, make-environment-variables, make-ephemeron, make-ephemeron-hash, make-ephemeron-hashalw, make-ephemeron-hasheq, make-ephemeron-hasheqv, make-exn, make-exn:break, make-exn:break:hang-up, make-exn:break:terminate, make-exn:fail, make-exn:fail:contract, make-exn:fail:contract:arity, make-exn:fail:contract:blame, make-exn:fail:contract:continuation, make-exn:fail:contract:divide-by-zero, make-exn:fail:contract:non-fixnum-result, make-exn:fail:contract:variable, make-exn:fail:filesystem, make-exn:fail:filesystem:errno, make-exn:fail:filesystem:exists, make-exn:fail:filesystem:missing-module, make-exn:fail:filesystem:version, make-exn:fail:network, make-exn:fail:network:errno, make-exn:fail:object, make-exn:fail:out-of-memory, make-exn:fail:read, make-exn:fail:read:eof, make-exn:fail:read:non-char, make-exn:fail:syntax, make-exn:fail:syntax:missing-module, make-exn:fail:syntax:unbound, make-exn:fail:unsupported, make-exn:fail:user, make-file-or-directory-link, make-flat-contract, make-fsemaphore, make-generic, make-hash, make-hash-placeholder, make-hashalw, make-hashalw-placeholder, make-hasheq, make-hasheq-placeholder, make-hasheqv, make-hasheqv-placeholder, make-immutable-hash, make-immutable-hashalw, make-immutable-hasheq, make-immutable-hasheqv, make-impersonator-property, make-input-port, make-input-port/read-to-peek, make-inspector, make-interned-syntax-introducer, make-keyword-procedure, make-known-char-range-list, make-list, make-lock-file-name, make-log-receiver, make-logger, make-mixin-contract, make-none/c, make-output-port, make-parameter, make-parent-directory*, make-phantom-bytes, make-pipe, make-pipe-with-specials, make-placeholder, make-plumber, make-polar, make-portal-syntax, make-prefab-struct, make-primitive-class, make-proj-contract, make-pseudo-random-generator, make-reader-graph, make-readtable, make-rectangular, make-rename-transformer, make-resolved-module-path, make-security-guard, make-semaphore, make-set!-transformer, make-shared-bytes, make-sibling-inspector, make-special-comment, make-srcloc, make-string, make-struct-field-accessor, make-struct-field-mutator, make-struct-type, make-struct-type-property, make-syntax-delta-introducer, make-syntax-introducer, make-tentative-pretty-print-output-port, make-thread-cell, make-thread-group, make-vector, make-weak-box, make-weak-hash, make-weak-hashalw, make-weak-hasheq, make-weak-hasheqv, make-will-executor, map, match-equality-test, matches-arity-exactly?, max, mcar, mcdr, mcons, member, member-name-key-hash-code, member-name-key=?, member-name-key?, memf, memory-order-acquire, memory-order-release, memq, memv, memw, merge-input, method-in-interface?, min, mixin-contract, module->exports, module->imports, module->indirect-exports, module->language-info, module->namespace, module->realm, module-cache-clear!, module-compiled-cross-phase-persistent?, module-compiled-exports, module-compiled-imports, module-compiled-indirect-exports, module-compiled-language-info, module-compiled-name, module-compiled-realm, module-compiled-submodules, module-declared?, module-path-index-join, module-path-index-resolve, module-path-index-split, module-path-index-submodule, module-path-index?, module-path?, module-predefined?, module-provide-protected?, modulo, mpair?, mutable-set, mutable-setalw, mutable-seteq, mutable-seteqv, n->th, nack-guard-evt, namespace-anchor->empty-namespace, namespace-anchor->namespace, namespace-anchor?, namespace-attach-module, namespace-attach-module-declaration, namespace-base-phase, namespace-call-with-registry-lock, namespace-mapped-symbols, namespace-module-identifier, namespace-module-registry, namespace-require, namespace-require/constant, namespace-require/copy, namespace-require/expansion-time, namespace-set-variable-value!, namespace-symbol->identifier, namespace-syntax-introduce, namespace-undefine-variable!, namespace-unprotect-module, namespace-variable-value, namespace?, nan?, natural-number/c, natural?, negate, negative-integer?, negative?, never-evt, newline, ninth, non-empty-string?, nonnegative-integer?, nonpositive-integer?, normal-case-path, normalize-arity, normalize-path, normalized-arity?, not, null, null?, number->string, number?, numerator, object\%, object->vector, object-info, object-interface, object-method-arity-includes?, object-name, object-or-false=?, object=-hash-code, object=?, object?, odd?, open-input-bytes, open-input-string, open-output-bytes, open-output-nowhere, open-output-string, order-of-magnitude, ormap, other-execute-bit, other-permission-bits, other-read-bit, other-write-bit, output-port?, pair?, parameter-procedure=?, parameter?, parameterization?, parse-command-line, partition, path->bytes, path->complete-path, path->directory-path, path->string, path-add-extension, path-add-suffix, path-convention-type, path-element->bytes, path-element->string, path-element?, path-for-some-system?, path-get-extension, path-has-extension?, path-list-string->path-list, path-only, path-replace-extension, path-replace-suffix, path-string?, path<?, path?, peek-byte, peek-byte-or-special, peek-bytes, peek-bytes!, peek-bytes-avail!, peek-bytes-avail!*, peek-bytes-avail!/enable-break, peek-char, peek-char-or-special, peek-string, peek-string!, permutations, phantom-bytes?, pi, pi.f, pipe-content-length, place-break, place-channel, place-channel-get, place-channel-put, place-channel-put/get, place-channel?, place-dead-evt, place-enabled?, place-kill, place-location?, place-message-allowed?, place-wait, place?, placeholder-get, placeholder-set!, placeholder?, plumber-add-flush!, plumber-flush-all, plumber-flush-handle-remove!, plumber-flush-handle?, plumber?, poll-guard-evt, port->list, port-closed-evt, port-closed?, port-commit-peeked, port-count-lines!, port-count-lines-enabled, port-counts-lines?, port-display-handler, port-file-identity, port-file-unlock, port-next-location, port-number?, port-print-handler, port-progress-evt, port-provides-progress-evts?, port-read-handler, port-try-file-lock?, port-waiting-peer?, port-write-handler, port-writes-atomic?, port-writes-special?, port?, portal-syntax-content, portal-syntax?, positive-integer?, positive?, predicate/c, prefab-key->struct-type, prefab-key?, prefab-struct-key, prefab-struct-type-key+field-count, preferences-lock-file-mode, pregexp, pregexp-quote, pregexp?, pretty-display, pretty-print, pretty-print-.-symbol-without-bars, pretty-print-abbreviate-read-macros, pretty-print-columns, pretty-print-current-style-table, pretty-print-depth, pretty-print-exact-as-decimal, pretty-print-extend-style-table, pretty-print-handler, pretty-print-newline, pretty-print-post-print-hook, pretty-print-pre-print-hook, pretty-print-print-hook, pretty-print-print-line, pretty-print-remap-stylable, pretty-print-show-inexactness, pretty-print-size-hook, pretty-print-style-table?, pretty-printing, pretty-write, primitive-closure?, primitive-result-arity, primitive?, print, print-as-expression, print-boolean-long-form, print-box, print-graph, print-hash-table, print-mpair-curly-braces, print-pair-curly-braces, print-reader-abbreviations, print-struct, print-syntax-width, print-unreadable, print-value-columns, print-vector-length, printable/c, printable<\%>, printf, println, procedure->method, procedure-arity, procedure-arity-includes?, procedure-arity-mask, procedure-arity?, procedure-closure-contents-eq?, procedure-extract-target, procedure-impersonator*?, procedure-keywords, procedure-realm, procedure-reduce-arity, procedure-reduce-arity-mask, procedure-reduce-keyword-arity, procedure-reduce-keyword-arity-mask, procedure-rename, procedure-result-arity, procedure-specialize, procedure-struct-type?, procedure?, processor-count, progress-evt?, promise-forced?, promise-running?, promise/name?, promise?, prop:arity-string, prop:arrow-contract, prop:arrow-contract-get-info, prop:arrow-contract?, prop:authentic, prop:blame, prop:chaperone-contract, prop:checked-procedure, prop:contract, prop:contracted, prop:custom-print-quotable, prop:custom-write, prop:dict, prop:equal+hash, prop:evt, prop:exn:missing-module, prop:exn:srclocs, prop:expansion-contexts, prop:flat-contract, prop:impersonator-of, prop:input-port, prop:liberal-define-context, prop:object-name, prop:orc-contract, prop:orc-contract-get-subcontracts, prop:orc-contract?, prop:output-port, prop:place-location, prop:procedure, prop:recursive-contract, prop:recursive-contract-unroll, prop:recursive-contract?, prop:rename-transformer, prop:sealed, prop:sequence, prop:set!-transformer, prop:stream, proper-subset?, pseudo-random-generator->vector, pseudo-random-generator-vector?, pseudo-random-generator?, put-preferences, putenv, quotient, quotient/remainder, radians->degrees, raise, raise-argument-error, raise-argument-error*, raise-arguments-error, raise-arguments-error*, raise-arity-error, raise-arity-error*, raise-arity-mask-error, raise-arity-mask-error*, raise-contract-error, raise-mismatch-error, raise-range-error, raise-range-error*, raise-result-arity-error, raise-result-arity-error*, raise-result-error, raise-result-error*, raise-type-error, raise-user-error, random, random-seed, rational?, rationalize, read, read-accept-bar-quote, read-accept-box, read-accept-compiled, read-accept-dot, read-accept-graph, read-accept-infix-dot, read-accept-lang, read-accept-quasiquote, read-accept-reader, read-byte, read-byte-or-special, read-bytes, read-bytes!, read-bytes-avail!, read-bytes-avail!*, read-bytes-avail!/enable-break, read-bytes-line, read-case-sensitive, read-cdot, read-char, read-char-or-special, read-curly-brace-as-paren, read-curly-brace-with-tag, read-decimal-as-inexact, read-eval-print-loop, read-installation-configuration-table, read-language, read-line, read-on-demand-source, read-single-flonum, read-square-bracket-as-paren, read-square-bracket-with-tag, read-string, read-string!, read-syntax, read-syntax-accept-graph, read-syntax/recursive, read/recursive, readtable-mapping, readtable?, real->decimal-string, real->double-flonum, real->floating-point-bytes, real->single-flonum, real-part, real?, reencode-input-port, reencode-output-port, regexp, regexp-match, regexp-match-exact?, regexp-match-peek, regexp-match-peek-immediate, regexp-match-peek-positions, regexp-match-peek-positions-immediate, regexp-match-peek-positions-immediate/end, regexp-match-peek-positions/end, regexp-match-positions, regexp-match-positions/end, regexp-match/end, regexp-match?, regexp-max-lookbehind, regexp-quote, regexp-replace, regexp-replace*, regexp-replace-quote, regexp-replaces, regexp-split, regexp-try-match, regexp?, regular-file-type-bits, relative-path?, remainder, remf, remf*, remove, remove*, remq, remq*, remv, remv*, remw, remw*, rename-contract, rename-file-or-directory, rename-transformer-target, rename-transformer?, replace-evt, reroot-path, resolve-path, resolved-module-path-name, resolved-module-path?, rest, reverse, round, second, seconds->date, security-guard?, semaphore-peek-evt, semaphore-peek-evt?, semaphore-post, semaphore-try-wait?, semaphore-wait, semaphore-wait/enable-break, semaphore?, sequence->list, sequence->stream, sequence-add-between, sequence-andmap, sequence-append, sequence-count, sequence-filter, sequence-fold, sequence-for-each, sequence-generate, sequence-generate*, sequence-length, sequence-map, sequence-ormap, sequence-ref, sequence-tail, sequence/c, sequence?, set, set!-transformer-procedure, set!-transformer?, set->list, set->stream, set-add, set-add!, set-box!, set-box*!, set-clear, set-clear!, set-copy, set-copy-clear, set-count, set-empty?, set-eq?, set-equal-always?, set-equal?, set-eqv?, set-first, set-for-each, set-group-id-bit, set-implements/c, set-implements?, set-intersect, set-intersect!, set-map, set-mcar!, set-mcdr!, set-member?, set-mutable?, set-phantom-bytes!, set-port-next-location!, set-remove, set-remove!, set-rest, set-subtract, set-subtract!, set-symmetric-difference, set-symmetric-difference!, set-union, set-union!, set-user-id-bit, set-weak?, set=?, set?, setalw, seteq, seteqv, seventh, sgn, sha1-bytes, sha224-bytes, sha256-bytes, shared-bytes, shell-execute, shrink-path-wrt, shuffle, simple-form-path, simplify-path, sin, single-flonum-available?, single-flonum?, sinh, sixth, skip-projection-wrapper?, sleep, socket-type-bits, some-system-path->string, special-comment-value, special-comment?, special-filter-input-port, split-at, split-at-right, split-common-prefix, split-path, splitf-at, splitf-at-right, sqr, sqrt, srcloc->string, srcloc-column, srcloc-line, srcloc-position, srcloc-source, srcloc-span, srcloc?, stencil-vector, stencil-vector-length, stencil-vector-mask, stencil-vector-mask-width, stencil-vector-ref, stencil-vector-set!, stencil-vector-update, stencil-vector?, sticky-bit, stop-after, stop-before, stream->list, stream-add-between, stream-andmap, stream-append, stream-count, stream-empty?, stream-filter, stream-first, stream-fold, stream-for-each, stream-force, stream-length, stream-map, stream-ormap, stream-ref, stream-rest, stream-tail, stream-take, stream/c, stream?, string, string->bytes/latin-1, string->bytes/locale, string->bytes/utf-8, string->immutable-string, string->keyword, string->list, string->number, string->path, string->path-element, string->some-system-path, string->symbol, string->uninterned-symbol, string->unreadable-symbol, string-append, string-append*, string-append-immutable, string-ci<=?, string-ci<?, string-ci=?, string-ci>=?, string-ci>?, string-contains?, string-copy, string-copy!, string-downcase, string-environment-variable-name?, string-fill!, string-foldcase, string-grapheme-count, string-grapheme-span, string-length, string-locale-ci<?, string-locale-ci=?, string-locale-ci>?, string-locale-downcase, string-locale-upcase, string-locale<?, string-locale=?, string-locale>?, string-no-nuls?, string-normalize-nfc, string-normalize-nfd, string-normalize-nfkc, string-normalize-nfkd, string-port?, string-prefix?, string-ref, string-set!, string-suffix?, string-titlecase, string-upcase, string-utf-8-length, string<=?, string<?, string=?, string>=?, string>?, string?, struct->vector, struct-accessor-procedure?, struct-constructor-procedure?, struct-info, struct-mutator-procedure?, struct-predicate-procedure?, struct-type-authentic?, struct-type-info, struct-type-make-constructor, struct-type-make-predicate, struct-type-property-accessor-procedure?, struct-type-property-predicate-procedure?, struct-type-property/c, struct-type-property?, struct-type-sealed?, struct-type?, struct:arity-at-least, struct:arrow-contract-info, struct:date, struct:date*, struct:exn, struct:exn:break, struct:exn:break:hang-up, struct:exn:break:terminate, struct:exn:fail, struct:exn:fail:contract, struct:exn:fail:contract:arity, struct:exn:fail:contract:blame, struct:exn:fail:contract:continuation, struct:exn:fail:contract:divide-by-zero, struct:exn:fail:contract:non-fixnum-result, struct:exn:fail:contract:variable, struct:exn:fail:filesystem, struct:exn:fail:filesystem:errno, struct:exn:fail:filesystem:exists, struct:exn:fail:filesystem:missing-module, struct:exn:fail:filesystem:version, struct:exn:fail:network, struct:exn:fail:network:errno, struct:exn:fail:object, struct:exn:fail:out-of-memory, struct:exn:fail:read, struct:exn:fail:read:eof, struct:exn:fail:read:non-char, struct:exn:fail:syntax, struct:exn:fail:syntax:missing-module, struct:exn:fail:syntax:unbound, struct:exn:fail:unsupported, struct:exn:fail:user, struct:srcloc, struct?, sub1, subbytes, subclass?, subclass?/c, subprocess, subprocess-group-enabled, subprocess-kill, subprocess-pid, subprocess-status, subprocess-wait, subprocess?, subset?, substring, suggest/c, symbol->string, symbol-interned?, symbol-unreadable?, symbol<?, symbol=?, symbol?, symbolic-link-type-bits, sync, sync/enable-break, sync/timeout, sync/timeout/enable-break, syntax->datum, syntax->list, syntax-arm, syntax-binding-set, syntax-binding-set->syntax, syntax-binding-set?, syntax-bound-phases, syntax-bound-symbols, syntax-column, syntax-debug-info, syntax-disarm, syntax-e, syntax-line, syntax-local-apply-transformer, syntax-local-bind-syntaxes, syntax-local-certifier, syntax-local-context, syntax-local-expand-expression, syntax-local-get-shadower, syntax-local-identifier-as-binding, syntax-local-introduce, syntax-local-lift-context, syntax-local-lift-expression, syntax-local-lift-module, syntax-local-lift-module-end-declaration, syntax-local-lift-provide, syntax-local-lift-require, syntax-local-lift-values-expression, syntax-local-make-definition-context, syntax-local-make-definition-context-introducer, syntax-local-make-delta-introducer, syntax-local-module-defined-identifiers, syntax-local-module-exports, syntax-local-module-interned-scope-symbols, syntax-local-module-required-identifiers, syntax-local-name, syntax-local-phase-level, syntax-local-submodules, syntax-local-transforming-module-provides?, syntax-local-value, syntax-local-value/immediate, syntax-original?, syntax-position, syntax-property, syntax-property-preserved?, syntax-property-remove, syntax-property-symbol-keys, syntax-protect, syntax-rearm, syntax-recertify, syntax-shift-phase-level, syntax-source, syntax-source-module, syntax-span, syntax-taint, syntax-tainted?, syntax-track-origin, syntax-transforming-module-expression?, syntax-transforming-with-lifts?, syntax-transforming?, syntax?, system-big-endian?, system-idle-evt, system-language+country, system-library-subpath, system-path-convention-type, system-type, tail-marks-match?, take, take-common-prefix, take-right, takef, takef-right, tan, tanh, tcp-abandon-port, tcp-accept, tcp-accept-evt, tcp-accept-ready?, tcp-accept/enable-break, tcp-addresses, tcp-close, tcp-connect, tcp-connect/enable-break, tcp-listen, tcp-listener?, tcp-port?, tentative-pretty-print-port-cancel, tentative-pretty-print-port-transfer, tenth, terminal-port?, the-unsupplied-arg, third, thread, thread-cell-ref, thread-cell-set!, thread-cell-values?, thread-cell?, thread-dead-evt, thread-dead?, thread-group?, thread-receive, thread-receive-evt, thread-resume, thread-resume-evt, thread-rewind-receive, thread-running?, thread-send, thread-suspend, thread-suspend-evt, thread-try-receive, thread-wait, thread/suspend-to-kill, thread?, time-apply, touch, true, truncate, udp-addresses, udp-bind!, udp-bound?, udp-close, udp-connect!, udp-connected?, udp-multicast-interface, udp-multicast-join-group!, udp-multicast-leave-group!, udp-multicast-loopback?, udp-multicast-set-interface!, udp-multicast-set-loopback!, udp-multicast-set-ttl!, udp-multicast-ttl, udp-open-socket, udp-receive!, udp-receive!*, udp-receive!-evt, udp-receive!/enable-break, udp-receive-ready-evt, udp-send, udp-send*, udp-send-evt, udp-send-ready-evt, udp-send-to, udp-send-to*, udp-send-to-evt, udp-send-to/enable-break, udp-send/enable-break, udp-set-receive-buffer-size!, udp-set-ttl!, udp-ttl, udp?, unbox, unbox*, uncaught-exception-handler, unit?, unquoted-printing-string, unquoted-printing-string-value, unquoted-printing-string?, unspecified-dom, unsupplied-arg?, use-collection-link-paths, use-compiled-file-check, use-compiled-file-paths, use-user-specific-search-paths, user-execute-bit, user-permission-bits, user-read-bit, user-write-bit, value-blame, value-contract, values, variable-reference->empty-namespace, variable-reference->module-base-phase, variable-reference->module-declaration-inspector, variable-reference->module-path-index, variable-reference->module-source, variable-reference->namespace, variable-reference->phase, variable-reference->resolved-module-path, variable-reference-constant?, variable-reference-from-unsafe?, variable-reference?, vector, vector*-append, vector*-copy, vector*-extend, vector*-length, vector*-ref, vector*-set!, vector*-set/copy, vector->immutable-vector, vector->list, vector->pseudo-random-generator, vector->pseudo-random-generator!, vector->values, vector-append, vector-argmax, vector-argmin, vector-cas!, vector-copy, vector-copy!, vector-count, vector-drop, vector-drop-right, vector-empty?, vector-extend, vector-fill!, vector-filter, vector-filter-not, vector-immutable, vector-length, vector-map, vector-map!, vector-member, vector-memq, vector-memv, vector-ref, vector-set!, vector-set*!, vector-set-performance-stats!, vector-set/copy, vector-split-at, vector-split-at-right, vector-take, vector-take-right, vector?, version, void, void?, weak-box-value, weak-box?, weak-set, weak-setalw, weak-seteq, weak-seteqv, will-execute, will-executor?, will-register, will-try-execute, with-input-from-bytes, with-input-from-string, with-output-to-bytes, with-output-to-string, would-be-future, wrap-evt, writable<\%>, write, write-byte, write-bytes, write-bytes-avail, write-bytes-avail*, write-bytes-avail-evt, write-bytes-avail/enable-break, write-char, write-special, write-special-avail*, write-special-evt, write-string, writeln, xor, zero?, check-tail-contract, for-clause-syntax-protect, legacy-match-expander?, match-...-nesting, match-expander?, prop:legacy-match-expander, prop:match-expander, syntax-local-match-introduce, syntax-pattern-variable?, \#\%app, \#\%datum, \#\%declare, \#\%expression, \#\%module-begin, \#\%plain-app, \#\%plain-lambda, \#\%plain-module-begin, \#\%printing-module-begin, \#\%provide, \#\%require, \#\%stratified-body, \#\%top, \#\%top-interaction, \#\%variable-reference, ->, ->*, ->*m, ->d, ->dm, ->i, ->m, ..., :do-in, <=/c, =/c, ==, =>, >=/c, _, absent, abstract, add-between, all-defined-out, all-from-out, and, and/c, any, any/c, apply, arity-at-least, arrow-contract-info, augment, augment*, augment-final, augment-final*, augride, augride*, bad-number-of-results, begin, begin-for-syntax, begin0, between/c, blame-add-context, box-immutable/c, box/c, call-with-atomic-output-file, call-with-file-lock/timeout, call-with-input-file, call-with-input-file*, call-with-output-file, call-with-output-file*, case, case->, case->m, case-lambda, channel/c, char-in/c, check-duplicates, class, class*, class-field-accessor, class-field-mutator, class/c, class/derived, combine-in, combine-out, command-line, compound-unit, compound-unit/infer, cond, cons/c, cons/dc, continuation-mark-key/c, contract, contract-exercise, contract-in, contract-out, contract-pos/neg-doubling, contract-struct, contracted, copy-directory/files, current-contract-region, date, date*, define, define-compound-unit, define-compound-unit/infer, define-contract-struct, define-custom-hash-types, define-custom-set-types, define-for-syntax, define-local-member-name, define-logger, define-match-expander, define-member-name, define-module-boundary-contract, define-namespace-anchor, define-opt/c, define-sequence-syntax, define-serializable-class, define-serializable-class*, define-signature, define-signature-form, define-splicing-for-clause-syntax, define-struct, define-struct/contract, define-struct/derived, define-syntax, define-syntax-rule, define-syntaxes, define-unit, define-unit-binding, define-unit-from-context, define-unit/contract, define-unit/new-import-export, define-unit/s, define-values, define-values-for-export, define-values-for-syntax, define-values/invoke-unit, define-values/invoke-unit/infer, define/augment, define/augment-final, define/augride, define/contract, define/final-prop, define/match, define/overment, define/override, define/override-final, define/private, define/public, define/public-final, define/pubment, define/subexpression-pos-prop, define/subexpression-pos-prop/name, delay, delay/idle, delay/name, delay/strict, delay/sync, delay/thread, delete-directory/files, dict->list, dict-can-functional-set?, dict-can-remove-keys?, dict-clear, dict-clear!, dict-copy, dict-count, dict-empty?, dict-for-each, dict-has-key?, dict-implements/c, dict-implements?, dict-iterate-first, dict-iterate-key, dict-iterate-next, dict-iterate-value, dict-keys, dict-map, dict-map/copy, dict-mutable?, dict-ref, dict-ref!, dict-remove, dict-remove!, dict-set, dict-set!, dict-set*, dict-set*!, dict-update, dict-update!, dict-values, dict?, display-lines, display-lines-to-file, display-to-file, do, dynamic->*, dynamic-instantiate, dynamic-place, dynamic-place*, else, eof-evt, except, except-in, except-out, exn, exn:break, exn:break:hang-up, exn:break:terminate, exn:fail, exn:fail:contract, exn:fail:contract:arity, exn:fail:contract:blame, exn:fail:contract:continuation, exn:fail:contract:divide-by-zero, exn:fail:contract:non-fixnum-result, exn:fail:contract:variable, exn:fail:filesystem, exn:fail:filesystem:errno, exn:fail:filesystem:exists, exn:fail:filesystem:missing-module, exn:fail:filesystem:version, exn:fail:network, exn:fail:network:errno, exn:fail:object, exn:fail:out-of-memory, exn:fail:read, exn:fail:read:eof, exn:fail:read:non-char, exn:fail:syntax, exn:fail:syntax:missing-module, exn:fail:syntax:unbound, exn:fail:unsupported, exn:fail:user, export, extends, failure-cont, field, field-bound?, file, file->bytes, file->bytes-lines, file->lines, file->list, file->string, file->value, find-files, find-relative-path, first-or/c, flat-contract-with-explanation, flat-murec-contract, flat-rec-contract, for, for*, for*/and, for*/async, for*/first, for*/fold, for*/fold/derived, for*/foldr, for*/foldr/derived, for*/hash, for*/hashalw, for*/hasheq, for*/hasheqv, for*/last, for*/list, for*/list/concurrent, for*/lists, for*/mutable-set, for*/mutable-setalw, for*/mutable-seteq, for*/mutable-seteqv, for*/or, for*/product, for*/set, for*/setalw, for*/seteq, for*/seteqv, for*/stream, for*/sum, for*/vector, for*/weak-set, for*/weak-setalw, for*/weak-seteq, for*/weak-seteqv, for-label, for-meta, for-space, for-syntax, for-template, for/and, for/async, for/first, for/fold, for/fold/derived, for/foldr, for/foldr/derived, for/hash, for/hashalw, for/hasheq, for/hasheqv, for/last, for/list, for/list/concurrent, for/lists, for/mutable-set, for/mutable-setalw, for/mutable-seteq, for/mutable-seteqv, for/or, for/product, for/set, for/setalw, for/seteq, for/seteqv, for/stream, for/sum, for/vector, for/weak-set, for/weak-setalw, for/weak-seteq, for/weak-seteqv, gen:custom-write, gen:dict, gen:equal+hash, gen:equal-mode+hash, gen:set, gen:stream, generic, get-field, get-preference, hash-copy-clear, hash-map/copy, hash/c, hash/dc, if, implies, import, in-bytes, in-bytes-lines, in-dict, in-dict-keys, in-dict-values, in-directory, in-ephemeron-hash, in-ephemeron-hash-keys, in-ephemeron-hash-pairs, in-ephemeron-hash-values, in-hash, in-hash-keys, in-hash-pairs, in-hash-values, in-immutable-hash, in-immutable-hash-keys, in-immutable-hash-pairs, in-immutable-hash-values, in-immutable-set, in-inclusive-range, in-indexed, in-input-port-bytes, in-input-port-chars, in-lines, in-list, in-mlist, in-mutable-hash, in-mutable-hash-keys, in-mutable-hash-pairs, in-mutable-hash-values, in-mutable-set, in-naturals, in-port, in-producer, in-range, in-set, in-slice, in-stream, in-string, in-syntax, in-value, in-vector, in-weak-hash, in-weak-hash-keys, in-weak-hash-pairs, in-weak-hash-values, in-weak-set, include, include-at/relative-to, include-at/relative-to/reader, include/reader, inclusive-range, inherit, inherit-field, inherit/inner, inherit/super, init, init-depend, init-field, init-rest, initiate-sequence, inner, input-port-append, inspect, instanceof/c, instantiate, integer-in, interface, interface*, invariant-assertion, invoke-unit, invoke-unit/infer, lambda, lazy, let, let*, let*-values, let-syntax, let-syntaxes, let-values, let/cc, let/ec, letrec, letrec-syntax, letrec-syntaxes, letrec-syntaxes+values, letrec-values, lib, link, list*of, list/c, listof, local, local-require, log-debug, log-error, log-fatal, log-info, log-warning, make-custom-hash, make-custom-hash-types, make-custom-set, make-custom-set-types, make-handle-get-preference-locked, make-immutable-custom-hash, make-limited-input-port, make-mutable-custom-set, make-object, make-temporary-directory, make-temporary-directory*, make-temporary-file, make-temporary-file*, make-weak-custom-hash, make-weak-custom-set, match, match*, match*/derived, match-define, match-define-values, match-lambda, match-lambda*, match-lambda**, match-let, match-let*, match-let*-values, match-let-values, match-letrec, match-letrec-values, match/derived, match/values, member-name-key, mixin, module, module*, module+, mutable-treelist/c, nand, new, new-∀/c, new-∃/c, non-empty-listof, none/c, nor, not/c, object-contract, object/c, one-of/c, only, only-in, only-meta-in, only-space-in, open, open-input-file, open-input-output-file, open-output-file, opt/c, or, or/c, overment, overment*, override, override*, override-final, override-final*, parameter/c, parameterize, parameterize*, parameterize-break, parametric->/c, pathlist-closure, peek-bytes!-evt, peek-bytes-avail!-evt, peek-bytes-evt, peek-string!-evt, peek-string-evt, peeking-input-port, place, place*, place/context, planet, port->bytes, port->bytes-lines, port->lines, port->string, prefix, prefix-in, prefix-out, pretty-format, private, private*, procedure-arity-includes/c, process, process*, process*/ports, process/ports, promise/c, prompt-tag/c, prop:dict/contract, property/c, protect-out, provide, provide-signature-elements, provide/contract, public, public*, public-final, public-final*, pubment, pubment*, quasiquote, quasisyntax, quasisyntax/loc, quote, quote-syntax, quote-syntax/prune, raise-blame-error, raise-not-cons-blame-error, raise-syntax-error, range, read-bytes!-evt, read-bytes-avail!-evt, read-bytes-evt, read-bytes-line-evt, read-line-evt, read-string!-evt, read-string-evt, real-in, recontract-out, recursive-contract, regexp-match*, regexp-match-evt, regexp-match-peek-positions*, regexp-match-positions*, relative-in, relocate-input-port, relocate-output-port, remove-duplicates, rename, rename-in, rename-inner, rename-out, rename-super, require, send, send*, send+, send-generic, send/apply, send/keyword-apply, set!, set!-values, set-field!, set/c, shared, sort, srcloc, stream, stream*, stream-cons, stream-lazy, string-join, string-len/c, string-normalize-spaces, string-replace, string-split, string-trim, struct, struct*, struct-copy, struct-field-index, struct-guard/c, struct-out, struct/c, struct/contract, struct/ctc, struct/dc, struct/derived, submod, super, super-instantiate, super-make-object, super-new, symbols, syntax, syntax-binding-set-extend, syntax-case, syntax-case*, syntax-deserialize, syntax-id-rules, syntax-rules, syntax-serialize, syntax/c, syntax/loc, system, system*, system*/exit-code, system/exit-code, tag, this, this\%, thunk, thunk*, time, transplant-input-port, transplant-output-port, treelist/c, unconstrained-domain->, unit, unit-from-context, unit/c, unit/new-import-export, unit/s, unless, unquote, unquote-splicing, unsyntax, unsyntax-splicing, values/drop, vector-immutable/c, vector-immutableof, vector-sort, vector-sort!, vector/c, vectorof, when, with-continuation-mark, with-contract, with-contract-continuation-mark, with-handlers, with-handlers*, with-input-from-file, with-method, with-output-to-file, with-syntax, write-to-file, ~.a, ~.s, ~.v, ~?, ~@, ~a, ~e, ~r, ~s, ~v, λ, equal?, equal, eq
		},
		keywordstyle=\color{blue},
		classoffset=3,
		morekeywords={import, export},
		keywordstyle=\color{green},
		classoffset=4,
		morekeywords={fsm, make-dfa, sm-apply, sm-states, sm-sigma, sm-start, sm-finals, sm-rules, sm-gamma, sm-showtransitions, sm-graph, gen-state, make-ndpda, make-ndfa, gen-regexp-word, singleton-regexp, kleenestar-regexp, concat-regexp, union-regexp, make-cfg, make-rg, make-csg, grammar-derive, grammar-not-derive, check-accept?, check-reject?},
		keywordstyle=\color{pakistangreen},
		classoffset=5,
		morekeywords={check-equal?, check-expect, check-equal, check-pred},
		keywordstyle=\color{sblue},
		classoffset=0,
		alsoletter={',`,-,/,>,<,\#,\%,?,=,*},
		moredelim=**[is][\color{lightgray}]{<<@<<}{>>@>>},
		moredelim=**[is][\itshape\color{OliveGreen}]{<<;<<}{>>;>>},
	}

\definecolor{sblue}{rgb}{0.14, 0.16, 0.48}
\definecolor{regalia}{rgb}{0.32, 0.18, 0.5}
\definecolor{palatinatepurple}{rgb}{0.41, 0.16, 0.38}
\definecolor{pakistangreen}{rgb}{0.0, 0.4, 0.0}
\definecolor{darkorange}{rgb}{1.0, 0.55, 0.0}
\newcommand{\racket}{\texttt{Racket}}

\newcommand{\fsm}{\texttt{FSM}}
\newcommand{\regexp}{\texttt{regexp}}

\newcommand{\quot}{\texttt{\textquotesingle{}}}

\newcommand{\flatt}{\texttt{FLAT}}

\newcommand{\rbe}{\texttt{RBE}}
\newcommand{\rbes}{\texttt{RBE}s}
\newcommand{\ndfa}{\texttt{ndfa}}
\newcommand{\ep}{\texttt{$\epsilon$}}

\definecolor{ForestGreen}{RGB}{34,139,34}
\definecolor{darkgreen}{rgb}{0.01, 0.75, 0.24}
\definecolor{deepsaffron}{rgb}{1.0, 0.6, 0.2}
\definecolor{celestialblue}{rgb}{0.29, 0.59, 0.82}

\title{A Design Recipe and Recipe-Based Errors \\ for Regular Expressions}

\author{Marco T. Moraz\'{a}n
\institute{Seton Hall University}
\email{morazanm@shu.edu}
\and
Shamil Dzhatdoyev
\institute{Axoni, USA}
\email{shamild11@gmail.com}
\and
Josephine Des Rosiers
\institute{Penguin Random House}
\email{josieadesrosiers@gmail.com}
\and
Tijana Mini\'{c}
\institute{University of Washington}
\email{tminic@uw.edu}
\and
Andr\'{e}s M. Garced
\institute{Seton Hall University}
\email{maldona2@shu.edu}
\and
David Anthony K. Fields
\institute{Seton Hall University}
\email{fieldsda@shu.edu}
}

\def\titlerunning{A Design Recipe and Recipe-Based Errors for Regular Expressions}
\def\authorrunning{Moraz\'{a}n, Dzhatdoyev, Des Rosiers, Mini\'{c}, Garced, Fields}

\begin{document}

\maketitle 

\begin{abstract}
This article presents a novel framework to provide Formal Languages and Automata Theory students design support for the development of regular expressions. This framework includes a design recipe for regular expressions and a customized error messaging system. The error messaging system produces recipe-based errors that include the step of the design recipe not successfully completed. Furthermore, the error messages follow the established practices of being concise, succinct, jargon-free, and nonprescriptive. In addition, a shorthand syntax developed for writing unit tests is described. The in-class use of the design recipe is illustrated, two debugging sessions using the described system are discussed, and the implementation of the error messaging system is briefly sketched.
\end{abstract}

\section{Introduction}
\label{intro}

In a Formal Languages and Automata Theory (\flatt{}) course, students are taught that the value of an expression can be a language. This introduction commonly starts with regular expressions (\regexp{}s) \cite{Hopcroft,Lewis,LinzRodger,PBFLAT,Rich,Sipser,Sudkamp,Gaudioso}. In essence, a regular expression is an algorithm description to generate words in a regular language using union, concatenation, and Kleene star operations over a given alphabet and empty \cite{Morazan-regexprs}. Computer Science students ought to become familiar with regular expressions given the important role they play in many software applications. For instance, regular expressions are used to search text for a given pattern \cite{Clarke,Knuth,Thompson}, to search for cancer cells \cite{Subramanian}, and to perform lexical analysis \cite{Appel,Johnson}.

Most textbooks discuss the development of \regexp{}s in a nonsystematic manner. The prevalent approach is to present a formal definition and some examples (e.g., \cite{Linz,LinzRodger,Martin,Rich,Sipser}). The implicit assumption is that students will learn how to define regular expressions without any further guidance nor a systematic development methodology.  Some approaches go further and suggest exploiting the inductive definition of regular expressions for their development. Hopcroft et al., for example, follow this approach to write \regexp{}s bottom-up (i.e., starting with expressions for singletons) \cite{Hopcroft}. Moraz\'{a}n proposes a top-down functional-programming-based approach, using \fsm{} \cite{fsm}, to implement regular expressions \cite{PBFLAT}, but falls short of outlining a systematic development process. 

Regardless of the approach taken, as any \flatt{} instructor knows, it is sometimes difficult to discern what students are trying to communicate with the \regexp{}s they develop. In our experience, there are at least two major reasons for these phenomena. The first is that many students use an unguided exploration approach to define a \regexp{}. More often than not, such students cannot articulate the idea behind a \regexp{} they develop and, if the expression is correct, they stumbled onto it by luck. The second is that many students improperly attempt to use \regexp{} constructors. This signals that beginning \flatt{} students, as in other areas of Computer Science \cite{Keuning,Race}, need immediate feedback on their designs. Instructors, of course, cannot sit with every student and offer feedback as they solve exercises. Offering immediate feedback, therefore, is only practical when using a programming-based approach. The most common medium for this type of feedback is error messages. To be useful, however, error messages must be clear \cite{Algaraibeh,Becker1,Becker4,Denny,Heemskerk,Traver}, concise \cite{Becker1,Denny}, not break the abstraction barrier expected by students \cite{Becker1}, provide context \cite{Becker1,Heemskerk,Wrenn}, use a positive tone \cite{Becker1}, avoid redundant information \cite{Becker1}, and use logical argumentation \cite{Barik2,Becker1}.

This article describes our efforts to address the two major reasons for the phenomena outlined above by extending the design- and programming-based \flatt{} curriculum put forth in the textbook \emph{Programming-Based Formal Languages and Automata Theory--Design, Implement, Validate, and Prove} \cite{PBFLAT}. We present a novel design recipe for programming regular expressions. Design recipes were first developed by Felleisen et al. to help instruct newcomers to programming \cite{HtDP2} and later extended by Moraz\'{a}n for a 2-semester introduction to programming \cite{drecipe,APS,APD}. The presented design recipe for regular expressions aims to help students develop regular expressions in a systematic fashion. We also present the error messaging system developed to support this design-based pedagogy. This error-messaging system extends to regular expressions prior similar work done for state machine design \cite{fsm-errors2}. At the heart of this system are \emph{recipe-based errors}\footnote{This term was coined by Rose Bohrer from the Worcester Polytechnic Institute.} (\rbes{}) that include a reference to the step of the design recipe not successfully completed. The article is organized as follows. \Cref{fsm} outlines \regexp{}s in \fsm{}. \Cref{dr} presents the novel design recipe for regular expressions and illustrates its use. \Cref{errors} describes the subset of the error messaging system that protects constructors for \regexp{}s and outlines a debugging session. \Cref{tests} describes the subset of the error messaging system that incorporates unit testing and outlines a debugging session. \Cref{impl} briefly discusses the error messaging system's implementation. \Cref{rw} contrasts with related work. Finally, \Cref{concls} presents concluding remarks and directions for future work.

\section{Regular Expressions in \fsm}
\label{fsm}

\fsm{} (\texttt{F}unctional \texttt{S}tate \texttt{M}achines) \cite{fsm}, is a domain-specific language, embedded in \racket{} \cite{RacketGuide}, for the \flatt{} classroom. It allows programmers to easily define instances of \regexp{}s. The goal is to allow students to design, implement, debug, and gain confidence in their regular expressions before submitting for grading or developing a proof.

The standard 6 varieties of \regexp{}s are supported: null, empty, singleton, union, concatenation, and Kleene star. It is explained to students that a \regexp{} defines a language and that \regexp{}s may be composed to build more complex languages.

\fsm{} provides an inductive definition of \regexp{}s based on the following 6 constructors:
\begin{alltt}
     A \regexp{} is either: 
       0.       \textcolor{pakistangreen}{null-regexp}:  \(\rightarrow\) regexp
       1.      \textcolor{pakistangreen}{empty-regexp}:  \(\rightarrow\) regexp
       2.  \textcolor{pakistangreen}{singleton-regexp}: string \(\rightarrow\) regexp
       3.      \textcolor{pakistangreen}{union-regexp}: \regexp{} \regexp{} \(\rightarrow\) regexp
       4.     \textcolor{pakistangreen}{concat-regexp}: \regexp{} \regexp{} \(\rightarrow\) regexp
       5. \textcolor{pakistangreen}{kleenestar-regexp}: \regexp{} \(\rightarrow\) regexp
\end{alltt}
The first three are base cases and the last three are inductive cases. Students are reminded that this recursive definition is useful, because all self-references are to smaller \regexp{}s. For instance, \texttt{r$_{\texttt{1}}$} is a smaller regular expression than \texttt{(\textcolor{pakistangreen}{union-regexp} r$_{\texttt{1}}$ r$_{\texttt{2}}$)} or \texttt{(\textcolor{pakistangreen}{kleenestar-regexp} r$_{\texttt{1}}$)}. The string provided to \textcolor{pakistangreen}{\texttt{singleton-regexp}} must be of length 1 and correspond to a letter in the Roman alphabet or one of: \texttt{\$}, \texttt{\&}, \texttt{!}, \texttt{*}.

There are observers to extract subcomponents. For example, \texttt{\textcolor{pakistangreen}{singleton-regexp-a}} extracts the string used to build a singleton \regexp{} and \textcolor{pakistangreen}{\texttt{concat-regexp-r2}} extracts the second regular expression used to build a concatenation \regexp{}. There are also predicates to determine the type of a regular expression. For instance, \texttt{\textcolor{pakistangreen}{union-regexp?}} determines if a given argument is a union \regexp{}. Finally, \texttt{\textcolor{pakistangreen}{gen-regexp-word}} nondeterministically generates a word in a given \regexp{}'s language.

Based on the inductive definition, students are taught to construct regular expressions using structural recursion:
\begin{enumerate}
  \item If a nonrecursive \regexp{} is needed, define it directly using the appropriate (base) constructor
  \item If a recursive \regexp{} is needed, determine the needed type as follows:
    \begin{itemize}
      \item If there is a choice among sublanguages, specify the choices and build a union \regexp{}
      \item If an amalgamation of words from different sublanguages is needed, specify the sublanguages and build a concatenation \regexp{}
      \item If an amalgamation of words from the same language is needed, specify the needed language and build a Kleene star \regexp{}
    \end{itemize}
\end{enumerate}
It is emphasized to students that if a recursive \regexp{} is needed, then it is written assuming that regular expressions for any sublanguage exist. The regular expressions for the sublanguages, of course, must also be written, albeit later. This assumes a top-down approach to programming regular expressions. In our experience, a bottom-up approach (e.g., first defining singleton \regexp{}s for a language's alphabet elements) only works well for the simplest languages.

\begin{figure}[t!]
\begin{lstlisting}[language=racket,escapechar=\%]
  #lang fsm
  (define A (singleton-regexp "a")) 
  ;; word -> Boolean
  ;; Purpose: Determine if the given word is in L(A)
  (define (in-A? w) (equal%\textcolor{blue}{?}% w (list %\quot{}%a)))
  (check-equal%\textcolor{sblue}{?}% (in-A? %\quot{}$\epsilon$%) #f)            (check-equal%\textcolor{sblue}{?}% (in-A? %\quot{}%()) #f)
  (check-equal%\textcolor{sblue}{?}% (in-A? %\quot{}%(a a)) #f)
  (check-pred in-A? (gen-regexp-word A)) (check-pred in-A? (gen-regexp-word A))
     
  (define B (singleton-regexp "b"))
  ;; word -> Boolean
  ;; Purpose: Determine if the given word is in L(B)
  (define (in-B? w) (equal%\textcolor{blue}{?}% w (list %\quot{}%b)))
  (check-equal%\textcolor{sblue}{?}% (in-B? %\quot{}$\epsilon$%) #f)         (check-equal%\textcolor{sblue}{?}% (in-B? %\quot{}%()) #f)
  (check-equal%\textcolor{sblue}{?}% (in-B? %\quot{}%(b b b)) #f)
  (check-pred in-B? (gen-regexp-word B))     
  (check-pred in-B? (gen-regexp-word B))
     
  (define B*  (kleenestar-regexp B))
  ;; word -> Boolean
  ;; Purpose: Determine if given word is in b*
  (define (in-B*? w) (or (eq? w %\quot{}$\epsilon$%) (andmap (lambda (s) (eq? s %\quot{}%b)) w)))
  (check-equal%\textcolor{sblue}{?}% (in-B*? %\quot{}%(a)) #f)     (check-equal%\textcolor{sblue}{?}% (in-B*? %\quot{}%()) #t)
  (check-equal%\textcolor{sblue}{?}% (in-B*? %\quot{}%(a a a a a a)) #f)
  (check-pred in-B*? (gen-regexp-word B*))
  (check-pred in-B*? (gen-regexp-word B*))
     
  (define B*A (concat-regexp B* A))
  ;; word -> Boolean
  ;; Purpose: Determine if given word is in b*a
  (define (in-B*A? w)
    (and (eq? (last w) %\quot{}%a) (andmap (lambda (a) (eq? a %\quot{}%b)) (drop-right w 1))))
  (check-equal%\textcolor{sblue}{?}% (in-B*A? %\quot{}$\epsilon$%) #f)             (check-equal%\textcolor{sblue}{?}% (in-B*A? %\quot{}%(a a)) #f)
  (check-equal%\textcolor{sblue}{?}% (in-B*A? %\quot{}%(b b a b b)) #f)   (check-equal%\textcolor{sblue}{?}% (in-B*A? %\quot{}%(b b)) #f)
  (check-pred in-B*A? (gen-regexp-word B*A))
  (check-pred in-B*A? (gen-regexp-word B*A))
\end{lstlisting}
\caption{An \fsm{} program for \texttt{L} = \texttt{b}$^{\texttt{*}}$\texttt{a} using \regexp{}s.} \label{sample-regexp}
\end{figure} 

To illustrate programming in \fsm{}, consider the following language over the alphabet \texttt{$\Sigma$=\{a, b\}}:
\begin{alltt}
     L = \{w|w \textrm{starts with an arbitrary number of} \texttt{b}\textrm{s} \textrm{and ends with an} a\}
\end{alltt}
The \fsm{} program is displayed in \Cref{sample-regexp}. The development of this program may start by writing a \regexp{} for \texttt{L}. The words in \texttt{L} start with an arbitrary number of \texttt{b}s repeated (i.e., 0 or more), thus, suggesting defining an auxiliary \regexp{}, \texttt{B*}, to generate them. In addition, the words in \texttt{L} end with an \texttt{a}, thus, suggesting defining another auxiliary \regexp{}, \texttt{A}, to generate it. To create a word in \texttt{L}, a word generated by \texttt{B*} needs to be concatenated with a word generated by \texttt{A}. The \regexp{} for \texttt{L} may be defined as on line 28 in \Cref{sample-regexp}. Unit tests are written using \texttt{RackUnit} \cite{RackUnit} and \fsm{}'s \textcolor{pakistangreen}{\texttt{gen-regexp-word}} as displayed on lines 33--36. A predicate must be written to determine if a word is in \texttt{L} as displayed on lines 29--32. Finally, the auxiliary \regexp{}s are developed in a similar fashion and, in the interest of brevity, their development is omitted.

We observe that for some readers of the code, the tests may not fully document the program. Such readers make two observations. The first is that no concrete words are directly used to illustrate words in \texttt{L(B*A)}. That is, there are no concrete examples of words in the language. The second is that \texttt{B*A} is not used to test words that are not in its language (i.e., only the auxiliary predicate is used for this). At this point in the course, however, students have not been exposed to finite-state machines and their equivalence with \regexp{}s. Therefore, to ask them to write such tests would be outright confusing and rather cumbersome (i.e., transforming \texttt{B*A} to a nondeterministic finite-state machine). In part, this motivates our work on integrating testing into the error messaging system in order to automate the use of concrete words and hide the cumbersome details of testing such words.

\section{A Design Recipe for Regular Expressions}
\label{dr}

As in many Computer Science courses, a significant number of students tend to develop their programs using an unguided exploration strategy, which results in designs that are difficult to understand and/or finding a program that works by luck. This presents four significant problems. First, an unclear design makes the job of establishing the correctness of a regular expression significantly more difficult. Second, grading becomes complicated given that the instructor cannot discern what the student is thinking. Third, it is discouraging for many female students because it tends to favor ``hacking'' skills (i.e., long hours of unguided/random exploration), which leads to feelings that success is due to luck and failure is due to lack of ability \cite{Cantwell}. This has an impact on Computer Science as a whole, because such beliefs/perceptions undermine women's confidence more than men \cite{Christensen} and undermine women's interest in Computer Science \cite{Ehrlinger}. Fourth, it reinforces the fixed mindset some students have about their programming abilities (i.e. either they are innately good at it or they are innately not good at it, frequently depending on whether or not their unguided explorations are usually successful). Some students, for example, believe that programming intelligence is manifested by being able to complete a project or an assignment on their own without any type of assistance and do not believe that programming intelligence grows with effort and persistence \cite{ORourke}.

To address these concerns, we present students with a design recipe for regular expressions. A design recipe is a series of steps, each with a concrete outcome, that guides a student from a problem statement to a solution. Design recipes provide scaffolding for students engaged in problem solving and program implementation. The design recipe for regular expressions helps students develop a clear design, which makes it significantly easier to argue correctness. It also serves as a \emph{lingua franca} between students and instructors to discuss problem solving and programming using \regexp{}s. Thus, making it much easier for the instructor to discern what a student is thinking and grade their programs. Finally, the design recipe for regular expressions discourages unguided/random exploration, reduces the role of luck in finding a solution, and associates failure with not following its steps rather than with a lack of ability. This, in turn, ought to encourage the belief that programming intelligence grows with effort and persistence.

\subsection{Design Recipe}

\begin{figure}[t!]
\begin{enumerate}
     \item Identify the input alphabet, pick a name for the regular expression, and describe the language
     
     \item Identify the sublanguages and outline how to compose them
     
     \item Define a predicate to determine if a word is in the target language
     
     \item Write unit tests
     
     \item Define the regular expression
     
     \item Run the tests and, if necessary, debug by revisiting the previous steps
     \item Prove that the regular expression is correct
\end{enumerate}
\caption{Design Recipe for Regular Expressions.} \label{dr-regexps}
\end{figure}

The design recipe for regular expressions is displayed in \Cref{dr-regexps}. Let \texttt{L} denote the language for the \regexp{} being designed. The first step requires identifying \texttt{L}'s alphabet, choosing a descriptive name for the \regexp{}, and describing \texttt{L}. For the second step, a design idea is outlined. It encourages using a divide and conquer strategy by identifying sublanguages that are needed to build words in \texttt{L} and by outlining how words in these sublanguages need to be composed. A word in a sublanguage is generated by an auxiliary regular expression. This step is akin to identifying the need for auxiliary functions and determining how to compose them when programming in a general-purpose programming language. Students that have traversed a design-based programming curriculum (e.g., using \cite{APS} and \cite{APD} or using \cite{HtDP2}) find these steps immediately familiar. Students that have not traversed such a curriculum may find this step more challenging and require an introduction to problem decomposition.

The development of unit tests is done in two steps. For Step 3, a predicate is developed to determine if a given word is a member of \texttt{L}. This predicate must also be designed and include, for example, a signature, a purpose statement, and unit tests. For Step 4, unit tests are developed. Testing must be thorough covering both words in and not in \texttt{L}. For words in \texttt{L} tests are written using \fsm{}'s \textcolor{pakistangreen}{\texttt{gen-regexp-word}} and the predicate developed for Step 3. For such tests the use of \textcolor{sblue}{\texttt{check-pred}} \cite{RackUnit} is encouraged (not required), given that it simplifies test writing. For words not in \texttt{L}, tests are written by calling the developed predicate with inlined concrete words (usually using \textcolor{sblue}{\texttt{check-equal?}} \cite{RackUnit}). 

For Step 5, a regular expression is documented and defined using the results for 1--4. For Step 6, students run the tests and, if necessary, debug by revisiting the steps of the design recipe. It is common for there to be bugs related to misuse of \regexp{} constructors used to satisfy Step 5. To help address these, our recipe-based errors include the step of the design recipe not successfully completed as described in \Cref{errors}. 

Finally, to satisfy Step 7 a correctness argument is developed. This is done by assuming that the auxiliary regular expressions are correct and arguing that the composition of the words generated by the auxiliary regular expressions results in a word that is in \texttt{L}. 

A natural question that arises is: Does it suffice to just give students this design recipe? Based on our experience using design recipes with beginners, we expect the answer to be an unequivocal no. Students need to be led to appreciate the value of the design recipe. Our solution is to present students with a problem that requires careful problem decomposition (i.e., design) to find a solution. Otherwise, it is all too tempting for students to take short cuts by skipping steps.

\subsection{Illustrative Example}

\begin{figure}[t!]
%\centering
\begin{alltt}
  DNA sequences consist of an arbitrary number of four nucleotide bases:
      adenine (a)  guanine (g)  cytosine(c)  thymine (t)
  Certain genetic disorders, such as Huntington's disease, are characterized
  by containing the subsequence \textrm{cag} repeated two or more times in a row.

  To help test programs written to detect this disorder, it is useful to 
  generate DNA sequences that contain such a subsequence. Design and implement 
  a regular expression to generate DNA sequences with \textrm{cag} repeated two or more 
  times in a row.
\end{alltt}
\caption{A Computational Biology DNA-based problem.} \label{sample-problem}
\end{figure}

To illustrate development using the design recipe for regular expressions, students are presented with a DNA-based Computational Biology problem related to detecting Huntington's disease \cite{CAG}. The simplified problem statement is displayed in \Cref{sample-problem}. Such a problem highlights the importance of regular expressions beyond the \flatt{} classroom. Making such a connection to a real-world problem is especially important to help dispel perceptions that Computer Science is narrowly focused on programming, which has been identified as a factor that discourages women \cite{Fisher,Spieler,Vekiri}. 

To satisfy Step 1, the alphabet is identified as \texttt{\quot{}(a g c t)} and the name chosen is \texttt{DISORDER-DNA}. To satisfy Step 2, the following sublanguages are identified:
\begin{description}[leftmargin=!,labelwidth=\widthof{\bfseries CAG++},labelindent=\parindent]

\item[DNA] An arbitrary DNA strand

\item[CAG++] A DNA strand of two or more \texttt{cag} repetitions

\end{description}
To complete Step 2, a word in \texttt{L(DISORDER-DNA)} is generated by concatenating a word in \texttt{DNA}, a word in \texttt{CAG++}, and a word in \texttt{DNA}. 

For a word to be in \texttt{L(DISORDER-DNA)}, it must be a valid DNA strand, not the empty word, and contain at least two consecutive instances of \texttt{\quot{}(c a g)}. This leads to the development of the following predicate to satisfy Step 3:
\begin{lstlisting}[language=racket,escapechar=\%,numbers=none]
;;word -> boolean
;;Purpose: Determines if the given word is in L(DISORDER-DNA)
(define (in-DISORDER-DNA? a-word)
  (and (in-DNA? a-word)
       (not (eq? a-word %\quot{}$\epsilon$%))
       (>= (length a-word) 6)
       (or (in-CAG++? (take a-word 6))
           (in-DISORDER-DNA? (rest a-word)))))
\end{lstlisting}
The auxiliary predicates, \texttt{in-DNA?} and \texttt{in-CAG++?}, are implemented as part of the design process, respectively, for the \texttt{DNA} and \texttt{CAG++} \regexp{}s. We note that \texttt{in-DISORDER-DNA?}'s design is straightforward using structural recursion. 

\begin{figure}[t!]
\begin{lstlisting}[language=racket,escapechar=\%]
     ;; Sigma = {c a g}
     ;; Language: DNA strands containing only two or more consecutive instances of cag
     ;; Sublanguages:  CAG and CAG+
     (define CAG++ (concat-regexp CAG CAG+))

     ;;word -> Boolean   ;;Purpose: Determines if the given word is a part of L(CAG++)
     (define (in-CAG++? a-word)
       (and (not (eq? a-word EMP)) (>= (length a-word) 6) (in-CAG+? a-word)))

     (check-equal? (in-CAG++? EMP) #f)
     (check-equal? (in-CAG++? %\quot{}%(c a g)) #f)
     (check-equal? (in-CAG++? %\quot{}%(c c g t a)) #f)
     (check-pred in-CAG++? (gen-regexp-word CAG++))
     (check-pred in-CAG++? (gen-regexp-word CAG++))
     
     ;; sigma = {a g c t}
     ;; Language: DNA strands containing two or more consecutive instances of cag
     ;; Sublanguages: DNA, CAG++
     (define DISORDER-DNA (concat-regexp DNA (concat-regexp CAG++ DNA)))  
     
     ;;word -> boolean
     ;;Purpose: Determines if the given word is in L(DISORDER-DNA)
     (define (in-DISORDER-DNA? a-word)
       (and (in-DNA? a-word)
            (not (eq? a-word %\quot{}$\epsilon$%))
            (%\textcolor{blue}{>=}% (length a-word) 6)
            (or (in-CAG++? (take a-word 6))
                (in-DISORDER-DNA? (rest a-word)))))

     (check-equal? (in-DISORDER-DNA? %\quot{}$\epsilon$%) #f)
     (check-equal? (in-DISORDER-DNA? %\quot{}%(a c g t)) #f)
     (check-equal? (in-DISORDER-DNA? %\quot{}%(c a t c c a a)) #f)
     (check-pred in-DISORDER-DNA? (gen-regexp-word DISORDER-DNA))
     (check-pred in-DISORDER-DNA? (gen-regexp-word DISORDER-DNA))
\end{lstlisting}
\caption{The implementation of \texttt{CAG++} and \texttt{DISORDER-DNA}.} \label{dna-regexps}
\end{figure}

To satisfy Step 5, a \regexp{} using the name chosen in Step 1 is defined. Auxiliary \regexp{}s for the sublanguages defined in Step 2 are composed as outlined in the same step. For our example, we define the \regexp{} as follows:
\begin{lstlisting}[language=racket,escapechar=\%,numbers=none]
(define DISORDER-DNA (concat-regexp DNA (concat-regexp CAG++ DNA)))
\end{lstlisting}
The resulting (partial) program for \texttt{DISORDER-DNA} is displayed on lines 16--34 in \Cref{dna-regexps}. Once the auxiliary \regexp{}s are designed and implemented Step 6 may be satisfied.

For Step 7, the correctness argument assumes that \texttt{CAG++} generates a DNA strand of two or more \texttt{cag} repetitions and that \texttt{DNA} generates an arbitrary DNA strand. This means that \texttt{DISORDER-DNA} concatenates 3 DNA strands: the first is an arbitrary DNA strand (which may or may not contain two or more \texttt{cag} repetitions in a row), the second is a DNA strand that only contains two or more repetitions of \texttt{cag}, and the third is an arbitrary DNA strand (which also may or may not contain two or more \texttt{cag} repetitions in a row). Therefore, \texttt{DISORDER-DNA} generates an arbitrary DNA strand that has at least one sequence of \texttt{cag} repetitions in a row of length greater than or equal to two. The reader can appreciate that this reasoning is well within the grasp of any student that has taken a course in Discrete Mathematics or Logic.

The auxiliary \regexp{}s, \texttt{CAG++} and \texttt{DNA}, are designed and implemented in a similar fashion, including any further auxiliary functions that may be needed. For instance, following the steps of the design recipe for \texttt{CAG++} yields the (partial) program on lines 1--14 in \Cref{dna-regexps}. The argument for correctness follows in a similar fashion: assume \texttt{CAG} and \texttt{CAG+} correctly generate words and argue that \texttt{CAG++} correctly generates words. In the interest of brevity, we do not discuss the remaining auxiliary \regexp{}s and predicates. For interested readers, \Cref{A1} lists the remaining \regexp{} definitions and \Cref{A2} lists the remaining predicates.

\section{The Error Messaging System and Constructors}
\label{errors}

\subsection{General Description}

A common source of frustration for students is misuse of \regexp{} constructors. Students describe such errors as ``silly'' once they understand them. Such a description may sound innocent, but it is not uncommon for these ``silly'' errors to evoke feelings that revolve around lack of talent. This is a clear indication that students, especially those first starting with \fsm{} \regexp{}s, need a robust error messaging system that helps them understand the reasons behind errors. To address this problem, we have extended the work done on \rbes{} for state machines \cite{fsm-errors2} to regular expressions. \rbes{} reinforce our design-based methodology by encouraging students to reason in terms of design steps. Each error message includes the step of the design recipe that has not been successfully completed along with a concise explanation of the error. By including a design recipe step in an error message, the search space for resolving the error is reduced and student attention is focused on debugging using the steps of the design recipe. %It is important to note that \rbes{} are not prescriptive given that, as others have noted \cite{Marceau1}, it is impossible to know what the student has in mind.

\rbes{} are designed using the following principles:
\begin{enumerate}
  \item Use jargon-free vocabulary \cite{Becker1,Denny,Leinonen}
  \item Use vocabulary familiar to students from classroom lectures and the textbook \cite{Becker1,Denny,Marceau1,Wrenn}
  \item Use a positive non-blaming tone \cite{Becker1}
  \item Use clear sentences \cite{Denny,Guo}
  \item Be specific: State the constructor parameter and given value for which the error is thrown \cite{Becker1,Denny}
  \item Concretely state the reason the value provided is in error \cite{Kohn,Traver,Shneiderman}
  \item Only report one error at a time \cite{Flowers,Nienaltowski}
  \item Only contain information relevant to the reported error \cite{Becker1,Wrenn}
  \item Prescriptive solutions are not offered \cite{Becker1,Marceau1,McCall}
\end{enumerate}
Items 1--4 are related to the composition of the error messages based on guidelines found in the literature on error messages. The goal is make sure error messages are neither cryptic nor intimidating. To this end, errors messages are phrased with language that ought to be familiar to students. In our experience and as noted by others \cite{Denny,Guo}, item 4 is especially important for students whose native language is not English.

Items 5--9 are related to the content of error messages that is specific for \fsm{} \regexp{}s. The goal is to help students understand the reason behind the error message in a logical manner without misleading them on how to solve the problem. To this end, the argument for which an incorrect value is provided, the provided value, and the reason the provided value is in error are integrated into the error message. In addition, only one error is reported at a time. This is done for two reasons. The first is to not overwhelm students with multiple, possibly unrelated, errors at once. The second is to help them focus on the resolution of the first error detected \cite{Becker2,Becker3,Munson}. Finally, the content of the error messages only contain information relevant to the error reported and does not offer a solution to make sure students are not led down an unfruitful path to resolve the error. 

\subsection{Debugging Session}
\label{debug}

\begin{figure}[t!]
\begin{lstlisting}[language=racket,escapechar=\%]
     #lang fsm

     (define ODDA
       (let%\textcolor{blue}{*}% [(A              (singleton-regexp "a"))
              (B              (singleton-regexp "b"))
              (BSTAR          (kleenestar-regexp %\quot{}%b))
              (BSTAR_A        (concat-regexp BSTAR "a"))
              (A_BSTAR_A      (concat-regexp A BSTAR_A))
              (B-OR-A_BSTAR_A (union-regexp  B A_BSTAR_A))
              (EVENASTAR      (kleenestar-regexp B-OR-A_BSTAR_A))
              (A-EVENASTAR    (concat-regexp "a" EVENASTAR))]
         (concat-regexp BSTAR A-EVENASTAR)))

     (define (in-ODDA? w)
       (let [(as (filter (lambda} (s) (eq? s %\quot{}%a)) w))]
         (odd? (length as))))
         
     (define (not-in-ODDA? w) (not (in-ODDA? w)))

     (check-pred not-in-ODDA? %\quot{}%())
     (check-pred not-in-ODDA? %\quot{}%(a a a a a a))
     (check-pred not-in-ODDA? %\quot{}%(a b b a))
     (check-pred not-in-ODDA? %\quot{}%(a b a a b a b))
     (check-pred in-ODDA? (gen-regexp-word ODDA))
     (check-pred in-ODDA? (gen-regexp-word ODDA))
     (check-pred in-ODDA? (gen-regexp-word ODDA))
     (check-pred in-ODDA? (gen-regexp-word ODDA))
     (check-pred in-ODDA? (gen-regexp-word ODDA))
\end{lstlisting}
\caption{A buggy \regexp{} for words with an odd number of \texttt{a}s.} \label{buggy-regexp}
%\Description{A buggy regular expression.}
\end{figure}

To illustrate a student debugging session, consider the incorrect program displayed in \Cref{buggy-regexp}. The goal is to implement a regular expression for all words with an odd number of \texttt{a}s over the alphabet \texttt{\{a, b\}}. The student has successfully corrected \fsm{} syntax errors, written a predicate to determine if a word has an odd number of \texttt{a}s, and written unit tests. The only bugs that remain are related to misusing \regexp{} constructors.

Upon evaluating the program, an error is thrown highlighting the following expression in the program:
\begin{lstlisting}[language=racket,mathescape=true,escapechar=\%,numbers=none]
(kleenestar-regexp %\quot{}%b)
\end{lstlisting}
The obtained error message is:
\begin{alltt}
    \textcolor{red}{Step five of the design recipe for regular expressions has not been
    successfully completed. The argument to kleenestar-regexp must be a  
    regular expression, but found: b}
\end{alltt}
Observe that the error message specifies the design step not successfully completed, the value provided, the argument for which the error is thrown, and the reason for the detected error without prescribing a solution. The student, however, only casually reads the error message and changes the highlighted expression to:
\begin{lstlisting}[language=racket,mathescape=true,escapechar=\%,numbers=none]
(kleenestar-regexp "b")
\end{lstlisting}
Upon reevaluating the program, the error thrown highlights the revised expression. The error message displayed is:
\begin{alltt}
    \textcolor{red}{Step five of the design recipe for regular expressions has not been 
    successfully completed. The argument to kleenestar-regexp must be a 
    regular expression, but found: "b"}
\end{alltt}
In essence, the students sees the same error for a different reason and this prompts reading the error message more carefully. The buggy expression is updated to:
\begin{lstlisting}[language=racket,escapechar=\%,numbers=none]
(kleenestar-regexp B)
\end{lstlisting}
Upon running the program again, an error is thrown and the highlighted expression is:
\begin{lstlisting}[language=racket,escapechar=\%,numbers=none]
(concat-regexp BSTAR "a")
\end{lstlisting}
The error message displayed reads:
\begin{alltt}
    \textcolor{red}{Step five of the design recipe for regular expressions has not been 
    successfully completed. The second argument to concat-regexp must be a 
    regular expression, but found: "a"}
\end{alltt}
Observe that the error message specifies the design step not successfully completed, the value in error, and the argument in error. Thus, carefully focusing the student's attention. The student changes \texttt{"a"} to \texttt{A}. The revaluation of the program highlights the following expression as containing an error:
\begin{lstlisting}[language=racket,escapechar=\%,numbers=none]
(concat-regexp "a" EVENASTAR)
\end{lstlisting}
The displayed error message is:
\begin{alltt}
    \textcolor{red}{Step five of the design recipe for regular expressions has not been 
    successfully completed. The first argument to concat-regexp must be a 
    regular expression, but found: "a"}
\end{alltt}
Observe that, once again, the error message attempts to focus the student's attention. This time the problem is with the first argument for the concatenation \regexp{} constructor. Upon changing \texttt{"a"} to \texttt{A}, the student runs the program again. No errors are thrown, the tests pass, and the student can be cautiously optimistic that the program works.

\section{The Error Messaging System and Tests}
\label{tests}

\subsection{General Description}

Despite having a working program for \texttt{ODDA} (i.e., the result of debugging the code in \Cref{buggy-regexp}), there are several testing features that are undesirable:
\begin{itemize}
  \item[$\circ$] Two predicates are written to test words in and words not in the language
  
  \item[$\circ$] There is a great deal of repeated code
  
  \item[$\circ$] There are no tests using the regular expression on literal words
\end{itemize}
The first is problematic, because, despite being trivial, it is monotonous work. The second is problematic, because repetitive tasks tend to lead to carelessness and, as a consequence, bugs. The third is problematic, because it does not fully document the program, given that there are no examples of concrete words in the \regexp{}'s language. 

To address these shortcomings, testing is integrated into the error messaging system. The testing components are provided as optional (keyword) parameters to \regexp{} constructors and provide a shorthand notation for testing. Making these optional arguments to a constructor allows for quick design prototyping and experimentation. The programmer may provide any subset of the following optional testing arguments to a regular expression constructor:
\begin{itemize}
  \item[$\circ$] The alphabet of the regular expression
  
  \item[$\circ$] A predicate to test if a generated word is in the \regexp{}'s language (i.e., the result of Step 3 of the design recipe)
      
  \item[$\circ$] The number of tests to run when a predicate is provided
  
  \item[$\circ$] A list of words that ought to be generated by the regular expression
  
  \item[$\circ$] A list of words that should not be generated by the regular expression
\end{itemize}
The alphabet, when provided and upon successful construction of the \regexp{}, is used to check that all words generated using the constructed \regexp{} and in the testing lists only contain alphabet elements, and that singletons are constructed with an alphabet element. The predicate, when provided, is used to eliminate the need to write tests, like those on lines 24--28 in \Cref{buggy-regexp}, using \texttt{gen-regexp-word}. The number of tests allows the programmer to customize how many tests to run using words produced by \texttt{gen-regexp-word}. The list of words that ought to be generated addresses the need to provide concrete examples of words in the language. The list of words that should not be generated addresses the need to provide concrete examples of words not in the language. Although the need to write repetitive testing code is eliminated, students may still write tests using the facilities provided by \texttt{RackUnit} \cite{RackUnit}. 

\begin{figure}[t!]
\begin{lstlisting}[language=racket,escapechar=\%]
  (define ODDA
    (let%\textcolor{blue}{*}% [(A (singleton-regexp "a"))
           (B (singleton-regexp "b"))
           (BSTAR          (kleenestar-regexp B))
           (BSTAR_A        (concat-regexp BSTAR A))
           (A_BSTAR_A      (concat-regexp A BSTAR_A))
           (B-OR-A_BSTAR_A (union-regexp B A_BSTAR_A))
           (EVENASTAR      (kleenestar-regexp B-OR-A_BSTAR_A))
           (A-EVENASTAR    (concat-regexp A EVENASTAR))]
      (concat-regexp BSTAR
                     A-EVENASTAR
                     #:sigma %\quot{}%(a b)
                     #:pred (lambda (w)
                              (let [(as (filter (lambda (s) (eq? s %\quot{}%a)) w))]
                                (odd? (length as))))
                     #:gen-cases   5
                     #:in-lang     %\quot{}%((a) (a a a) (b b a b b) (b a a b b b a a a))
                     #:not-in-lang %\quot{}%(() (a a a a) (b b b) (b a b a)))))
\end{lstlisting}
\caption{Simplified code for \texttt{ODDA} from \Cref{debug}.} \label{simplified-regexp}
\end{figure}

To illustrate the simplification achieved, consider the code obtained for the debugged \texttt{ODDA} from the previous section displayed in \Cref{simplified-regexp}. The 4 optional arguments specify, \{a b\}, the language's alphabet, the predicate to determine if a word is in the \regexp{}'s language (equivalent to \texttt{in-ODDA?} from \Cref{buggy-regexp}), and lists of words, \texttt{in-lang} and \texttt{not-in-lang}, that ought to be, respectively, accepted and rejected. Observe that there is no need for the programmer to provide a predicate to determine if a word ought to be rejected nor to write repetitive code for unit testing. For each of the locally defined regular expressions, the programmer may also provide optional testing arguments. We have omitted them here in the interest of brevity and in the interest of readability. Nonetheless, this is a feature that traditional unit testing does not provide: one cannot write (nor run) unit tests for local definitions.

\subsection{Debugging Session}

The integration of testing does much more than provide shorthand notation for unit tests. It also reinforces design by producing \rbes{}. For students, the debugging search space is reduced by encouraging them to focus on a precise design recipe step. The \rbes{} produced are described as follows for Steps 1, 3, and 6 of the design recipe:
\begin{description}
  \item[Step 1] The argument for the input alphabet:
    \begin{itemize}
      \item[$\circ$] Is not a list
      
      \item[$\circ$] Is a list, but contains invalid elements 
      
      \item[$\circ$] Contains repetitions
      
      \item[$\circ$] the regular expression generates words containing elements not in the input alphabet
    \end{itemize}
      
  \item[Step 3] The argument for the predicate:
    \begin{itemize}
      \item[$\circ$] Produces a non-Boolean value
      
      \item[$\circ$] Does not hold when applied to a word generated using the regular expression
    \end{itemize}
  
  \item[Step 6] The arguments for:
  
    \begin{itemize}
      \item[$\circ$] The lists of concrete words to test:
        \begin{itemize}
          \item[$\ast$] Contain words with elements not in the \regexp{}'s alphabet
      
          \item[$\ast$] Contain words that are rejected that ought to be accepted
      
          \item[$\ast$] Contain words that are accepted that ought to be rejected
        \end{itemize}
      
      \item[$\circ$] The number of test words to generate for use with the given predicate:
        \begin{itemize}
          \item[$\ast$] Is not a positive integer
      
          \item[$\ast$] Is a number, but no predicate is provided (i.e., it is a dependent type; it depends on a predicate being provided)   
        \end{itemize}
    \end{itemize}
\end{description}

\begin{figure}[t!]
\begin{lstlisting}[language=racket,escapechar=\%]
     (define ab*Uba* 
      (let%\textcolor{blue}{*}% [(A       (singleton-regexp "a"))
             (B       (singleton-regexp "b"))
             (ASTAR   (kleenestar-regexp A))
             (BSTAR   (kleenestar-regexp B))
             (A-BSTAR (concat-regexp A BSTAR))
             (B-ASTAR (concat-regexp B ASTAR))
             (in-ab*Uba* 
               (lambda (w)
                 (and (not (empty? w))
                      (or (and (eq? (first w) 1) 
                               (andmap (lambda (s) (eq? s %\quot{}%b)) (rest w)))
                          (and (eq? (first w) %\quot{}%b) 
                               (andmap (lambda (s) (eq? s %\quot{}%a)) (rest w)))))))]
        (union-regexp 
          A-BSTAR
          B-ASTAR
          #:pred        in-ab*Uba*?
          #:gen-cases   %\quot{}%a
          #:sigma       %\quot{}%(a b)
          #:in-lang     %\quot{}%((a) (b) (a b b v) (b b) (b a a a a))
          #:not-in-lang %\quot{}%(() (a a a) (a b) (b a a a a b) (a b b b a a a b a)))))
\end{lstlisting}
\caption{Proposed regular expression for \texttt{L} = \texttt{ab$^{\texttt{*}}$ $\cup$ ba$^{\texttt{*}}$}.} \label{testing-regexp}
\end{figure}

To illustrate a debugging session, consider the proposed \regexp{} for \texttt{L} = \texttt{ab$^{\texttt{*}}$ $\cup$ ba$^{\texttt{*}}$} displayed in \Cref{testing-regexp}. Upon running the program the following \rbe{} is produced:
\begin{alltt}
     \textcolor{red}{Step six of the design recipe for regular expressions has not been
     successfully completed. The number of generated test cases to check
     with the predicate must be a positive integer, but found: a}
\end{alltt}
The student has provided the symbol \texttt{a} as the argument for the number of testing words to generate. Upon changing \texttt{a} to 3, the next \rbe{} generated is:
\begin{alltt}
     \textcolor{red}{Step three of the design recipe for regular expressions has not been 
     successfully completed. The given predicate does not hold for the 
     following words generated using the regexp: ((a b b b) (a b b b b b b b))}
\end{alltt}
The student can observe that the words ought to be accepted. Therefore, there must be a bug in the predicate. Upon redesigning the predicate, in the first branch of the \texttt{or}-expression, the 1 is changed to \texttt{\quot{}a}. Upon running the program, the following \rbe{} is generated:
\begin{alltt}
     \textcolor{red}{Step six of the design recipe for regular expressions has not been
     successfully completed. The following words to be generated by the regular 
     expression: ((a b b v)) contain characters not in the regular expression's 
     alphabet: (a b)}
\end{alltt}
Step six of the design recipe is not successfully completed, because the list of words that ought to be generated contains an invalid word. Upon fixing the mistyped letter by changing the \texttt{v} to a \texttt{b}, the following \rbe{} is generated:
\begin{alltt}
     \textcolor{red}{Step six of the design recipe for regular expressions has not been
     successfully completed. The following words are expected to be 
     generated by the constructed union-regexp but are not generated: ((b b))}
\end{alltt}
Step six of the design recipe is not successfully completed, because a word in the list of words that ought to be generated is not in the \regexp{}'s language. Upon moving \texttt{\quot{}(b b)} to the list of words not in the language, the following \rbe{} is generated:
\begin{alltt}
     \textcolor{red}{Step six of the design recipe for regular expressions has not been
     successfully completed. The following words are expected to not be 
     generated by the constructed union-regexp but are generated: ((a b))}
\end{alltt}
Step six of the design recipe is not successfully completed, because a word in the list of words that should not be generated is in the \regexp{}'s language. Upon moving \texttt{\quot{}(a b)} to the list of words that ought to be in the \regexp{}'s language, the \regexp{} is successfully built and validated.

\section{Implementation}
\label{impl}

This section sketches the contract-based implementation of the error messaging system for \regexp{}s. First, we provide a brief overview of \racket{} contracts. The goal is not to provide an exhaustive presentation. Instead, the goal is to present enough information to help readers unfamiliar with such contracts to navigate this section. Second, we illustrate \texttt{concat-regexp}'s contract-based implementation.

\subsection{Contracts in \racket{}}

Contracts are used to build more reliable software \cite{Meyer1}. They provide three desirable features to develop an error-messaging system. The first is that contracts help guarantee that the caller provides appropriate arguments and that the callee returns an appropriate value. In essence, contracts place assertions on the input and on the output, which are known as the precondition and the postcondition. The caller is obliged to satisfy the precondition for each argument and the callee is obliged to satisfy the postcondition. The second is that contracts can assign blame for errors more effectively than exceptions \cite{Lazarek,Plosch}. The third is that contracts allow a great deal of flexibility for customizing error messages. 

\racket{} provides a myriad of contract combinators \cite[Chapter 8]{RacketGuide}. Of these, this article makes use of the following three:
\begin{description}[leftmargin=!,labelwidth=\widthof{\bfseries and/c},labelindent=\parindent]
  \item[and/c] Produces a contract that takes as input an arbitrary number of contracts and accepts any value that satisfies all the given contracts \cite[Section~8.2]{RacketGuide}. The subcontracts are applied in the order given and the first to fail, if any, generates the needed \rbe{}.
  
  \item[-$\boldsymbol{>}$] Produces a contract for a function. It takes as input an arbitrary number of contracts such that the last contract is for the value returned and the rest are for the input values. The contracts for the inputs are applied, in the order provided, when the function is called and the first to fail, if any, generates an \rbe{}. The contract for the value returned is called after the function is applied to the arguments \cite[Section~7.2]{RacketGuide}.
      
  \item[-$\boldsymbol{>}$i] Like \textbf{-}$\boldsymbol{>}$, produces a contract for a function. It allows expressing dependencies between arguments and results \cite[Section~8.2]{RacketGuide}. Each argument may be named and, subsequently, used in subcontracts. It provides support for required, optional, and keyword parameters. For each argument, there is a stanza containing the argument's name, an optional list of dependencies, and a contract that must be satisfied.
\end{description}

\begin{figure}[t!]
\begin{lstlisting}[language=racket,escapechar=\%]
(define concat-regexp/c
 (->i ([r1 (valid-regexp/c ERR-ARG1)] [r2 (valid-regexp/c ERR-ARG2)])
      (#:sigma [sigma (r1 r2) 
        (and/c (is-a-list-regexp/c "regexp alphabet" "one")
               (valid-listof/c valid-alpha?     "lowercase alphabet letter" 
                               "input alphabet" #:rule "one" #:for-regexp? #t)
               (no-duplicates/c "sigma" "one" #:for-regexp? #t)
               (valid-regexp-sigma/c (make-unchecked-concat r1 r2)))]
       #:pred [pred (r1 r2 gen-cases) 
         (and/c valid-regexp-predicate/c 
                (predicate-passes/c (make-unchecked-concat r1 r2) gen-cases))]
       #:gen-cases [gen-cases valid-gen-cases-count/c]
       #:in-lang 
         [in-lang (r1 r2) 
          (let [(R (make-unchecked-concat r1 r2))] 
           (and/c (listof-words-regexp/c 
                    "words generated by the regular expression" 
                    "four")
                  (words-in-sigma/c "to be generated" R)
                  (regexp-input/c R %\quot{}%%\textcolor{black}{concat-regexp}% #t)))]
       #:not-in-lang 
         [not-in-lang (r1 r2) 
          (let [(R (make-unchecked-concat r1 r2))]
           (and/c (listof-words-regexp/c 
                    "words not generated by the regular expression" 
                    "four")
                  (words-in-sigma/c "not to be generated" R)
                  (regexp-input/c R %\quot{}%%\textcolor{black}{concat-regexp}% #f)))])
      [result concat-regexp?]))
\end{lstlisting}
\caption{Contract to build a concatenation regular expression.} \label{concat-contract}
\end{figure}

\subsection{Illustration of \regexp{} Contract}

To illustrate how \regexp{} contracts are implemented, the contract to build concatenation \regexp{}s is outlined. This constructor has two required arguments (i.e., for the two sub-\regexp{}s) and five optional keyword parameters (i.e., for the alphabet, the predicate, the number of test cases to use with the predicate, the test words in the language, and the test words not in the language).

The contract for \texttt{concat-regexp} is displayed in \Cref{concat-contract}. The required inputs, \texttt{r1} and \texttt{r2}, must be \regexp{}s as tested using the auxiliary contract \texttt{valid-regexp/c} on line 2. If either is not a \regexp{}, the \rbe{} generated indicates that Step 5 of the design recipe is not successfully completed using the provided strings (i.e., \texttt{ERR-ARG1} and \texttt{ERR-ARG2}).

The contracts for the optional arguments are displayed on lines 3--28 in \Cref{concat-contract}. These contracts are only applied if both given arguments are \regexp{}s. Except for the argument for the number of test cases to generate for the provided predicate (i.e., \texttt{\#:gen-cases}), all optional keyword arguments depend on the two input \regexp{}s, given that the concatenation regular expression must be built to perform testing. In addition to the two given regular expressions, \texttt{\#:pred} also depends on the argument provided for \texttt{\#:gen-cases}. These dependencies are bound to the variable names in parentheses. For example, on line 9 the arguments the contract for the predicate depends on are \texttt{r1} (declared on line 2), \texttt{r2} (declared on line 2), and \texttt{gen-cases} (declared on line 12). 

The contract for the alphabet argument (lines 3--8 in \Cref{concat-contract}) first tests if the given argument is a list (line 4), is a list of valid alphabet elements (lines 5--6), and contains no duplicates (line 7). If any of these contracts fail, the \rbe{} generated indicates that Step 1 of the design recipe is not successfully completed and the reason for the failure. If these contracts do not fail, then the concatenation regular expression is constructed (which is safe given that both given arguments are regular expressions) and the auxiliary contract \texttt{valid-regexp-sigma/c} tests if it contains any singleton regular expressions constructed using elements not in the given alphabet. If this contract fails, the \rbe{} generated indicates that Step 1 of the design recipe is not successfully completed and indicates the elements not contained in the given alphabet.

The contract for the predicate argument (lines 9--11  in \Cref{concat-contract}) first tests if the given argument is a predicate that is given as input a valid word using the auxiliary contract \texttt{valid-regexp-predicate/c} (line 10). This auxiliary contract is on a function that we define as follows:
\begin{alltt}
     (define valid-regexp-predicate/c
       (->  valid-word/c valid-regexp-predicate-result/c))
\end{alltt}
The contract on the result tests that it is a Boolean. If this contract fails, the \rbe{} generated indicates that Step 3 of the design recipe is unsuccessfully completed and the reason (i.e., invalid input argument for the word or a non-Boolean value is returned). If the predicate is valid, then the concatenation \regexp{} is constructed and used to test the number of cases specified by \texttt{gen-cases} (line 11). The auxiliary contract \texttt{predicate-passes/c} generates the necessary words using \texttt{gen-regexp-word} and applies the given predicate to them. If any words fail to satisfy the predicate then the generated \rbe{} indicates that Step 3 of the design recipe is unsuccessfully completed and lists the words for which the predicate fails.

The contract for the number of test words to generate determines if the provided argument is a positive integer using the auxiliary contract \texttt{valid-gen-cases-count/c} (line 12 in \Cref{concat-contract}). If this contract fails, the generated \rbe{} indicates that Step 6 of the design recipe is unsuccessfully completed, given that testing using the given predicate is not possible.

The contracts for the arguments denoting the concrete words that are in and not in the concatenation \regexp{}'s build, \texttt{R}, the concatenation regular expression (respectively, lines 14 and 19). They check that the given value is a list of words (respectively, lines 15 and 20), that the words are valid for \texttt{R}'s alphabet (respectively, lines 16 and 21), and that all provided concrete words are generated/not-generated (respectively, lines 17 and 22). To determine if all provided concrete words are generated/not-generated, \texttt{R} is converted to a nondeterministic finite-state machine (\ndfa{}) (using \fsm{}'s primitive \textcolor{pakistangreen}{\texttt{regexp->fsa}}). The \ndfa{} is used to determine, respectively, if the words are accepted or rejected. If the argument provided is not a list of words then the \rbe{} generated indicates that Step 4 of the design recipe is not successfully completed, given that no tests may be performed. If either of the other checks fail then the \rbe{} generated indicates that Step 6 of the design recipe is not successfully completed, given that tests fail, and provides the words that cause the test to fail.

\section{Related Work}
\label{rw}

There is a long and rich history of research on designing error messages for students albeit not specifically targeting regular expressions in a \flatt{} course. Universally, researchers agree that an error messaging system is among the most important tools that can be offered to students \cite{Becker1,Becker4,Denny,Hundhausen,Marceau2,Marceau1,Munson,Nienaltowski,Prather,Wrenn}. There is also universal agreement that error messages for students need to be comprehensible and not intimidating. There is some division on whether longer/enhanced error messages are more beneficial than shorter error messages. One study found that longer error messages are not more effective \cite{Nienaltowski} while another study was inconclusive \cite{Pettit}. Yet others, however, have found that longer messages help students understand and fix errors \cite{Prather}. On this front, we have opted to make \fsm{} error messages for regular expressions succinct/brief and concise/comprehensive. This design decision is made based on our experience with \fsm{}'s former error messaging system that listed multiple errors at once \cite{fsm-errors} and with \rbes{} for state machines \cite{fsm-errors2}. Listing multiple errors at once, makes error messages long and, in retrospect, not well-focused. In our experience, students were frustrated with the length of the messages and many did not read all the error messages. Instead, they opted to only read the last error message. As the literature suggests, this is an undesirable result given that the first error ought to be corrected first \cite{Becker2,Becker3,Munson}. In contrast, \rbes{} for state machines are short, only reporting one error at a time, and help focus students on design by referring to steps of the design recipe not successfully completed. Students have found such error messages useful, clear, and of better quality than those offered by other programming languages. For these reasons, error messaging for \regexp{}s emulates the approach taken for state machines.

The literature is also divided about offering prescriptive solutions in error messages. Proponents of offering prescriptive solutions argue that students can benefit from seeing what others have done to resolve a given error \cite{Hartmann}. The literature also warns that such solutions ought to only be suggested when there is a high degree of certainty that the solution is correct \cite{Kantorowitz}. A study by Hartmann \cite{Hartmann}, for example, found that only 47\% of the time proposed solutions were useful and 25\% of the time they were not useful. Detractors of offering prescriptive solutions argue that compilers and runtime systems cannot determine the intention of a programmer and, therefore, any prescriptive solution risks undermining this intention \cite{Becker1,Marceau1,McCall}. On this front, we agree with the detractors of prescriptive solutions. We have no reasonable means to predict the intention of \fsm{} programmers developing regular expressions and, therefore, restrict error messages to facts that can be verified by the programmer and that help explain why an error was detected.

\section{Concluding Remarks}
\label{concls}

This article presents novel design support to help students implement regular expressions in a domain-specific language, \fsm{}, for the Formal Languages and Automata Theory classroom. The support presented is a 7-step design recipe for the development of regular expressions that includes validation and verification steps. In addition, the features and implementation of a novel error messaging system built on recipe-based errors for regular expressions is presented. Recipe-based errors for regular expressions protect constructors and tightly-couple generated error messages with unsuccessfully completed design recipe steps. When an error is thrown the expression responsible for the error is highlighted and the generated error message is clear, informative, nonintimidating, succinct, concise, and nonprescriptive. The result is an error messaging system that provides students with the help they need to advance their implementations as illustrated by the presented debugging sessions. In addition, the error messaging system also includes features that provide a shorthand notation to write unit tests and couples test failures with unsuccessfully completed design recipe steps.

Future work includes conducting an empirical study to assess student perceptions of and disposition towards the presented error messaging system for regular expressions. In addition, future work includes implementing recipe-based errors for regular, context-free, and context-sensitive grammars in \fsm{}.

%%
%% The acknowledgments section is defined using the "acks" environment
%% (and NOT an unnumbered section). This ensures the proper
%% identification of the section in the article metadata, and the
%% consistent spelling of the heading.
%\begin{acks}

%\end{acks}

%%
%% The next two lines define the bibliography style to be used, and
%% the bibliography file.
\balance
\bibliographystyle{eptcs}
\bibliography{regexpr-errors}

%%
%% If your work has an appendix, this is the place to put it.
%%\appendix
\newpage

\appendix

The appendices list the remaining \regexp{}s definitions and the remaining predicate definitions for \texttt{DISORDER-DNA} from \Cref{dna-regexps}. In \Cref{A1}, the \regexp{} definitions are listed in \Cref{rem-defs}. In \Cref{A2}, the predicate definitions are listed in \Cref{rem-preds}, albeit without tests due to limitations on figure size.

\section{Remaining \regexp{}s}
\label{A1}

\begin{figure}[h!]
\begin{lstlisting}[language=racket,escapechar=\%]
;; Sigma = {a}  Language: (a)  Sublanguages: none
(define A (singleton-regexp "a"))

;; Sigma = {g}  Language: (g)  Sublanguages: none
(define G (singleton-regexp "g"))

;; Sigma = {c}  Language: (c)  Sublanguages: none
(define C (singleton-regexp "c"))

;; Sigma = {t}  Language: (t)  Sublanguages: none
(define T (singleton-regexp "t"))

;; Sigma = {a g c t}  Language: Any valid base nucleotide  Sublanguages: A, G, C, T
(define BASE (union-regexp A (union-regexp G (union-regexp C T))))

;; Sigma = {a g c t}  Language = any arbitrary DNA strand  Sublanguages: BASE
(define DNA (kleenestar-regexp BASE))

;; Sigma = {c a g}  Language = {(c a g)}  Sublanguages: C, A, G
(define CAG (concat-regexp C (concat-regexp A G)))

;; Sigma = {c a g}  Language = any DNA strand of 0 or more cag  Sublanguages: CAG
(define CAG* (kleenestar-regexp CAG))

;; Sigma = {c a g}  Language = any DNA strand of 1 or more cag  Sublanguages: CAG, CAG*
(define CAG+ (concat-regexp CAG CAG*))
\end{lstlisting}
\caption{Remaining \regexp{}s definitions.} \label{rem-defs}
\end{figure}

\newpage

\section{Predicates for Remaining \regexp{}s}
\label{A2}

\begin{figure}[h!]
\begin{lstlisting}[language=racket,escapechar=\%]
;; word -> boolean  Purpose: Determine if the given word is in L(A)
(define (in-A? a-word) (equal%\textcolor{blue}{?}% a-word (list %\quot{}%a)))

;; word -> boolean  Purpose: Determine if the given word is in L(G)
(define (in-G? a-word) (equal%\textcolor{blue}{?}% a-word (list %\quot{}%g)))

;; word -> boolean  Purpose: Determine if the given word is in L(C)
(define (in-C? a-word) (equal%\textcolor{blue}{?}% a-word (list %\quot{}%c)))

;; word -> boolean  Purpose: Determine if the given word is in L(T)
(define (in-T? a-word) (equal%\textcolor{blue}{?}% a-word (list %\quot{}%t)))

;; word -> boolean  Purpose: Determines if the given word is in L(BASE)
(define (in-BASE? a-word)
  (and (not (eq%\textcolor{blue}{?}% a-word %\quot{}$\epsilon$%))
       (or (in-A? a-word) (in-G? a-word) (in-C? a-word) (in-T? a-word))))

;; word -> Boolean  Purpose: Determines if the given word is apart of L(DNA)
(define (in-DNA? a-word)
  (or (equal%\textcolor{blue}{?}% a-word %\quot{}$\epsilon$%) (andmap (lambda (s) (in-BASE? (list s))) a-word)))

;; word -> boolean  Purpose: Determines if the given word is apart of L(CAG)
(define (in-CAG? a-word)
  (and (not (eq%\textcolor{blue}{?}% a-word %\quot{}$\epsilon$%))
       (= (length a-word) 3)
       (in-C? (list (first a-word)))
       (in-A? (list (second a-word)))
       (in-G? (list (third a-word)))))

;;word -> boolean  Purpose: Determines if the given word is in L(CAG*)
(define (in-CAG*? a-word)
  (or (equal%\textcolor{blue}{?}% a-word %\quot{}$\epsilon$%)
      (and (= (remainder (length a-word) 3) 0)
           (or (in-CAG? a-word)
               (and (in-CAG? (take a-word 3))
                    (in-CAG*? (drop a-word 3)))))))

;; word -> boolean  Purpose: Determines if the given is apart of L(CAG+)
(define (in-CAG+? a-word)
  (and (not (equal%\textcolor{blue}{?}% a-word %\quot{}$\epsilon$%)) (in-CAG*? a-word)))
\end{lstlisting}
\caption{Remaining predicates.} \label{rem-preds}
\end{figure}

\end{document}